\documentclass[12pt,a4paper]{article}

\usepackage[utf8]{inputenc}
\usepackage[T1]{fontenc}
\usepackage[margin=2.5cm]{geometry}
\usepackage[numbered]{bookmark}
\usepackage{mathtools}
\usepackage{ulem}
\usepackage{amssymb}
\usepackage{graphicx}
\usepackage[font=small,labelfont=bf]{caption}
\usepackage{authblk}
\usepackage{cite}
\usepackage{dsfont}
\usepackage{xcolor}

\usepackage[titles]{tocloft}

\numberwithin{equation}{section}

\newcommand{\be}{\begin{equation}}
	\newcommand{\ee}{\end{equation}}
\newcommand{\bea}{\begin{eqnarray}}
	\newcommand{\eea}{\end{eqnarray}}

\newcommand{\nocontentsline}[3]{}
\newcommand{\tocless}[2]{\bgroup\let\addcontentsline=\nocontentsline#1{#2}\egroup}

\title{ Circular strings in Kerr-$AdS_{5}$ black holes}
\author[a,b]{O. V. Geytota}
\author[a,c]{A. A. Golubtsova}
\author[a,d]{H.~Dimov}

\author[a,e]{Vu H. Nguyen}
\author[d,h]{R.~C.~Rashkov
	\footnote{Emails: \texttt{golubtsova, dimov  @theor.jinr.ru} and \texttt{h\_dimov,rash@phys.uni-sofia.bg}}}
{\footnotesize
\affil[a]{\textit{The Bogoliubov Laboratory of Theoretical Physics, JINR,}\authorcr\textit{141980 Dubna, Moscow region, Russia}\vspace{5pt} \vspace{3pt}} 
\affil[b]{\textit{Dubna State University}\authorcr\textit{Universitetskaya str.,141980 Dubna, Moscow region, Russia}\vspace{5pt} \vspace{3pt}}
\affil[c]{\textit{Steklov Mathematical Institute, Russian Academy of Sciences}\authorcr\textit{Gubkina str. 8, 119991 Moscow, Russia}\vspace{5pt} \vspace{3pt}} 
\affil[d]{\textit{Department of Physics, Sofia University,}\authorcr\textit{5 J. Bourchier Blvd., 1164 Sofia, Bulgaria}\vspace{5pt} \vspace{3pt}} 
\affil[e]{\textit{Institute of Physics, VAST,  10000 Hanoi, Vietnam}\vspace{5pt} \vspace{3pt}} 
\affil[h]{\textit{Institute for Theoretical Physics, Vienna University of Technology,}\authorcr\textit{Wiedner Hauptstr. 8--10, 1040 Vienna, Austria}}}

\date{}

\begin{document}
	\maketitle

	\begin{abstract}
	The quest for extension of holographic correspondence to the case of finite temperature naturally includes Kerr-AdS black holes and their field theory duals. We probe the five-dimensional Kerr-AdS space time  by pulsating strings. First we find particular pulsating string solutions and then semi-classically quantize the theory. For the string with large values of energy, we use the Bohr-Sommerfeld analysis to find the energy of the string as a function of a large quantum number. We obtain the wave function of the problem and thoroughly study the corrections to the energy, which according to the holographic dictionary are related to anomalous dimensions of certain operators in the dual gauge theory. The interpretation of results from holographic point of view is not straightforward since the dual theory is at finite temperature. Nevertheless, near or at conformal point the expressions can be thought of as the dispersion relations of stationary states.
	\end{abstract}
	\newpage
	\tableofcontents
	\newpage
	
	\section{Introduction}

The intensive development of the gauge/gravity correspondence has yielded useful toolkits for exploring non-perturbative regime of quantum field theories. 
The gauge/gravity duality states that all the physics in a d-dimensional conformal gauge theory at strong coupling can be described in terms of a gravitational theory in a $(d+1)$-dimensional spacetime with certain asymptotics.
In particular, 
the dynamics of the 4d $\mathcal{N} =4$ $SU(N)$ SYM in the strong coupling regime is equivalent to the dynamics of the classical IIB superstring theory on $AdS_{5}\times S^{5}$  in the weakly coupled regime.
This provides a dictionary between observables of the dual theories, which gives an opportunity to probe non-perturbative dynamics of gauge theories. An important class of observables includes various closed string configurations which energy spectrum can be related to anomalous dimensions of single-trace  local operators in SYM. 
Despite that computing the string spectrum even in asymptotically AdS backgrounds is an intricate problem, the integrability methods used within the holographic duality allowed to find and analyze many string configurations.

Addressing the important issues as strong coupling phenomena, it became of great interest to extend the holographic duality to  the case of finite temperature, which generically has less symmetries and phenomenologically more appropriate. Thermal observables contain a lot of information on dynamics of the system, however, they seems to be difficult to compute.
 The AdS/CFT correspondence allows to relate characteristics of black black holes in asymptotically AdS spacetimes  to observables of strongly coupled quantum systems.   The solutions are assumed to describe thermal states of the dual CFT with certain Hawking temperature. The use of the AdS/CFT correspondence appears to be a powerful method to investigate the thermal states of CFT. Indeed, the horizon is playing the role of a thermal background. This approach allows to include more dimensionless parameters in the theory making it very useful not only to collect important information for a higher dimensional theory  but also to study its holographic dual.
 Particularly, the AdS black holes with spherical horizon are dual to the thermal ensemble of $\mathcal{N} = 4$ SYM on $\mathbb{S}^{1}\times \mathbb{S}^{3}$, while a planar AdS black brane is dual to finite-temperature $\mathcal{N} = 4$ SYM on $\mathbb{S}^{1}\times \mathbb{R}^{3}$\cite{Witten, Witten2}. An intriguing suggestion has been made in \cite{HHT}, namely to consider a 5d Kerr-AdS black hole with a non-zero angular momentum as a gravity dual to "a rotating Einstein universe" on $\mathbb{R}\times \mathbb{S}^{3}$. 
 As a higher dimensional black hole the Kerr-$AdS_{5}$ black hole solution is characterized by two rotational parameters, related to the two parts of the angular momenta independently preserving.  These parameters can be associated to the rotation in different planes.  Note, that
 Kerr-AdS black holes  share with non-rotating AdS black holes a number of common interesting features including the Hawking-Page phase transition, scaling of the free energy  \cite{HReal} and found
 its application in a holographic description of a rotating quark-gluon plasma 
 \cite{NAS}-\cite{GGU2021}. At high temperatures the conformal symmetry is restored, so this description seems to be viable.
 
Certain thermal holographic observables can be found considering  string dynamics in the black hole backgrounds.
Circular closed strings in AdS black holes have been discussed in \cite{Petkou, Alishahiha:2002fi}. Instead of rotating strings in the pure AdS case, the strings in the black holes are orbiting in these backgrounds.
In particular, orbiting strings outside the 5d AdS-Schwarzschild black holes were interpreted  as states of large spins in the dual  thermal ensemble of $\mathcal{N} = 4$ SYM theory on $\mathbb{S}^{1}\times \mathbb{S}^{3}$. For the case of rotating AdS black holes  the thermodynamical stability of closed string in the 5d Kerr-AdS black holes was studied with respect to angular momentum leakage to the black hole. A generalization of a pulsating string \cite{min,Engquist:2003, Dimov:2004,Smedback:2004} to the 5d AdS-Schwarzschild background was suggested in \cite{Alishahiha:2002fi}. Following the Bohr-Sommerfeld quantization the energy of the string was computed, this energy can be associated with dispersion relations of the states in the $\mathcal{N} = 4$ SYM theory at finite temperature.
  
In this paper we study pulsating string configurations in the 5d Kerr-AdS black hole with equal rotational parameters. We construct a pulsating string solution in the black hole background.
We also compute the string energy reducing the string Nambu-Goto action to the mechanical Lagrangian and applying the Bohr-Sommerfeld analysis. The potential wells are related to the position of the outer black hole horizon and the boundary of the black hole spacetime.  We derive  the relation for the energy for the case of a small value of the rotational parameters.
 In the case of vanishing rotation this relation for the energy comes to that one obtained earlier in the work \cite{Alishahiha:2002fi}. Note,  that for the pure AdS case the energy of the string can be related to the anomalous dimensions of single trace operators in $\mathcal{N} = 4$ SYM theory. In the black hole case we  cannot establish this connection, since the notion of the anonymous dimension is defined in the conformal point. However,  we can think on its relevance to the dispersion relations of the states in the thermal ensemble of $\mathcal{N} = 4$ SYM theory on $\mathbb{S}^{1}\times \mathbb{S}^{3}$.
 We also perform a WKB approximation and obtain  the Schr\"odinger equation on the reduced subspace $y=const$.
 
 The paper is organized as follows. In Section ~\ref{Sect2} we give a review of the Kerr-$AdS_{5}$ black hole with equal rotational parameters.
 Section~\ref{Sect3} is devoted to a pulsating string solutions,  we present  solutions for a pulsating string in the Kerr-$AdS_{5}$.
 In Section \ref{Sect4} we consider the Bohr-Sommerfeld analysis  and the WKB approximation.  Finally, in Section \ref{Sect5} we conclude. In appendices we collect a number of useful relations for our computations.
 
\newpage	
	\section{5d Kerr-AdS black hole geometry}\label{Sect2}
	
	The  5d Kerr-AdS black hole solution was constructed in \cite{HHT} and describes a rotating black hole with an AdS asymptotic. The metric of the 5d Kerr-AdS black holes is characterized by two rotational parameters $a$, $b$, related to the Casimirs of $SU(2)\times SU(2)\simeq SO(4)$. In present paper  we focus on the case of equal rotational parameters $a=b$.
	The metric in the so-called AdS coordinates (static-at-infinity frame)  can be represented as follows
    \bea\label{GLKA}
	ds^{2}& =& -(1+ y^{2}\ell^{2})dT^{2} + y^{2}(d\Theta^{2} + \sin^{2}\Theta d\Phi^{2} + \cos^{2}\Theta d\Psi^{2}) \\
	&+& \frac{2M}{y^{2}\Xi^{3}}(dT - a \sin^{2}\Theta d\Phi - a\cos^{2}\Theta d\Psi)^{2}
	+\frac{y^{4}dy^{2}}{y^{4}(1 + y^{2}\ell^{2}) - \frac{2M}{\Xi^{2}}y^{2} + \frac{2Ma^{2}}{\Xi^{3}}},\nonumber
	\eea
where 
\be
\Xi = 1 - a^{2}\ell^2,
\ee
 $M$ is the mass of the black hole, $a$ is a rotational parameter and we use the Hopf coordinates to parametrize the metric on the sphere with $ 0\leq\Theta \leq \frac{\pi}{2}$,  $0\leq\Phi,\Psi\leq2\pi$.   Like ordinary Kerr solutions Kerr-AdS black holes have inner and outer horizons. We consider, that
 the holographic radial coordinate $y$ with values on the region $(y_{+}, +\infty)$, where $y_{+}$ is an outer horizon of the black hole and we reach an AdS-asymptotics  as $y$ goes to $+\infty$.

The position of the outer horizon in these coordinates is defined by a largest root of the equation {\footnote{We present the solutions to this equation in Appendix B.}}
\be
1 + y^{2}\ell^{2} - \frac{2M}{\Xi^{2}y^2} + \frac{2Ma^{2}}{\Xi^{3}y^4} =0.
\ee
Particularly, for the extremal 5d Kerr-AdS black hole the horizon is given by
	\be
	y^{2}_{\rm +, ext} = \frac{1}{4\Xi}\left[4a^{2}\ell^{2} - 1 + \sqrt{1 + 8a^{2}\ell^{2}}\right].
	\ee
	
From (\ref{GLKA})  it is easy to see that  with $y \to \infty$ the 4d  boundary of 5d Kerr-AdS black hole is 4d  $R\times S^{3}$ \cite{HReal, Gibbons:2004ai}
	\be
	ds^{2} = - dT^{2} + d\Theta^{2} +\sin^{2} \Theta d\Phi^{2} + \cos^{2}\Theta d \Psi^{2}.
	\ee

The Hawking temperature of the Kerr-AdS black hole is given by 
\be\label{hawkT}
T_{H} =  \frac{1}{2\pi}\left(\frac{2y_{+}(1+y^{2}_{+}\ell^2)}{(y^{2}_{+}+a^{2})} - \frac{1}{y_{+}}\right).
\ee
	The angular momentum and the angular velocity are given by
	\be
	J = \frac{\pi M a}{ \Xi^{3}}, \quad \Omega = \frac{a(1+ y^2_{+}\ell^2)}{y^{2}_{+}+a^{2}},
	\ee
	correspondingly.
Note, that there is a Hawking-Page phase transition in the Kerr-AdS black hole,  and the rotation  has an influence on it \cite{AGG2020}.

	Taking $M  =0$ in the 5d Kerr-AdS solution (\ref{GLKA}) we get 
	\be\label{Kerr-AdSab}
	ds^{2}= - (1+ y^{2}\ell^{2})dT^{2} + y^{2}(d\Theta^{2} + \sin^{2}\Theta d\Phi^{2} + \cos^{2}\Theta d\Psi^{2})^{2}
	+\frac{dy^{2}}{(1 + y^{2}\ell^{2})},
	\ee
	that is merely the global representation of the AdS metric.\\
	
	The 5d Kerr-AdS black hole  holographically is interpreted as a gravity dual of the 4d thermal $\mathcal{N} =4$ SYM theory on $\mathbb{R}\times \mathbb{S}^{3}$ (a thermal ensemble of  $\mathcal{N} =4$ SYM theory) at strong coupling \cite{HHT, HReal, Petkou, Alishahiha:2002fi}.  The temperature of the theory is defined by the Hawking temperature (\ref{hawkT}). Comparing to the pure AdS case, the energy of the string in the black hole background can not be identified to the scaling dimension, which is defined near the conformal points. However, it is still possible to relate these strings to single gauge-invariant states, so the string 	analysis can be linked to the dispersion relations of the stationary states of the thermal $\mathcal{N} =4$ SYM theory \cite{Alishahiha:2002fi}.


	\section{Exact solution of pulsating string in 5d Kerr-AdS background}
		\label{Sect3}
	
	The approach of pulsating strings for finding the anomalous dimensions of CFT operators starts with the work \cite{min} and further generalizations have been proposed in \cite{Engquist:2003, Dimov:2004, Smedback:2004}. Then a number of papers developing its application in the holographic systems appeared , see for instance \cite{Khan:2003sm, Arutyunov:2003za, Kruczenski:2004cn, Bobev:2004id, Park:2005kt, deVega:1994yz, Chen:2008qq, Dimov:2009rd, Arnaudov:2010by, Arnaudov:2010dk, Beccaria:2010zn, Giardino:2011jy, Pradhan:2013sja, Pradhan:2014zqa}. The purpose of this section is to obtain a pulsating string solutions in the Kerr-$AdS_5$ background. To this end we will construct the Polyakov action and find appropriate solutions.

The starting point is the Polyakov string action in the conformal gauge  written as follows
\begin{equation}\label{actP}
	S_{P} = - \frac{1}{4\pi \alpha'} \int d\tau d\sigma\{\sqrt{-h}h^{\alpha\beta}\partial_{\alpha}X^{M}\partial_{\beta}X^{N}G_{MN}\},
\end{equation}
where $h^{\alpha\beta} = \textrm{diag}(-1,1)$, $\alpha,\beta = 0,1$, $M,N=1,..,5$. \\
Using the notations in appendix A, the string Lagrangian in the Kerr-$AdS_5$ black hole (\ref{GLKA}) takes the following form
\begin{align}\label{LagrangianP}
	-4\pi\alpha'\mathcal{L}&=\,G_{TT}(T'^2-\dot{T}^2)+G_{yy}(y'^2-\dot{y}^2)+G_{\Theta\Theta}(\Theta'^2-\dot{\Theta}^2)\nonumber\\
	&+G_{\Phi\Phi}(\Phi'^2-\dot{\Phi}^2)+G_{\Psi\Psi}(\Psi'^2-\dot{\Psi}^2)\nonumber\\
	&+2G_{T\Phi}(T'\Phi'-\dot{T}\dot{\Phi})+2G_{T\Psi}(T'\Psi'-\dot{T}\dot{\Psi})+2G_{\Phi\Psi}(\Phi'\Psi'-\dot{\Phi}\dot{\Psi})\, ,
\end{align}
where we have used $ \,\dot{X}=\partial_{\tau} X  \, $ and $\,  X'=\partial_{\sigma} X\,$.\\
 The Virasoro constraints following from (\ref{actP}) are given by
\bea\label{Vir1}
\textrm{Vir1:}\quad &&\sum_{M,N}G_{MN}\left(\partial_{\tau}X^{M}\partial_{\tau}X^{N} + \partial_{\sigma}X^{M}\partial_{\sigma}X^{N}\right)= 0,\\ \label{Vir2}
\textrm{Vir2:}\quad&&\sum_{M,N}G_{MN}\partial_{\tau}X^{M}\partial_{\sigma}X^{N} = 0.
\eea
%
%
\subsection{Pulsating string configuration involving the  holographic direction "y"}
	\label{Sect31}
In view of our further considerations, we would like to obtain a classical pulsating string solution having dependence on the holographic direction "y". Through the following calculations we will show that such a solution exists. 
The ansatz for the pulsating string configuration, involving the $y$-direction, which is consistent with the equations of motion  is
\begin{equation}\label{y-ansatz}
	T=\kappa\,\tau,\qquad y=y(\tau),\qquad \Theta=\Theta^*\,=\,const,\qquad \Phi=\,m_{\phi}\sigma,\qquad \Psi=\,m_{\psi} \sigma\,.
\end{equation}
The Polyakov string lagrangian takes the form
\begin{equation}\label{P-Lagrangian}
	L_{P}\,\sim\,-G_{TT}\,\dot{T}^2 -G_{yy}\,\dot{y}^2 +G_{\Phi\Phi}\,m_{\phi}^2 +G_{\Psi\Psi}\,m_{\psi}^2 +2G_{\Phi\Psi}\,m_{\phi}\,m_{\psi}\,.
\end{equation}
Let's consider the Virasoro constraints (\ref{Vir2}) and (\ref{Vir2}) more precisely.  Substituting the string ansatz \eqref{y-ansatz}  into (\ref{Vir2}) can be rewritten as
\begin{equation}\label{Vir2new}
	Vir2: \qquad \kappa\, \frac{2aM}{y^2 \Xi^3}\left(m_{\phi}\,\sin^2\Theta +  m_{\psi}\, \cos^2\Theta  \right) =0\,.
\end{equation}
Since $y$ is not a constant, this condition fixes $\,\Theta=\Theta^*\,$ as follows 
\begin{equation}\label{Vir2 theta*}
	m_{\phi}\,\sin^2\Theta^* +  m_{\psi}\, \cos^2\Theta^*  =0\,\Rightarrow\, \tan^2 \Theta^* = \,-\,\frac{m_{\psi}}{m_{\phi}}\,>\,0,\, \textrm{sign}(m_{\phi})\,\neq\,\textrm{sign}(m_{\psi})\,.
\end{equation}
Substituting the ansatz \eqref{y-ansatz} in \eqref{Vir1}, we have
\begin{equation} \label{VIR1BAS}
	G_{TT}\kappa^2 + G_{yy}\,\dot{y}^2 +y^2\,\left( m_{\phi}^2\,\sin^2\Theta +  m_{\psi}^2\, \cos^2\Theta   \right) + \frac{2a^2 M}{y^2 \Xi^3}\left(m_{\phi}\,\sin^2\Theta +  m_{\psi}\, \cos^2\Theta  \right)^2 \,=\,0\,.
\end{equation}
Taking into account (\ref{Vir2new}), eq.\eqref{VIR1BAS} can be written as
\begin{equation}
	G_{TT}\kappa^2 + G_{yy}\,\dot{y}^2 +y^2\,\left( m_{\phi}^2\,\sin^2\Theta^* +  m_{\psi}^2\, \cos^2\Theta^*   \right)\,=\,0\,,
\end{equation}
or in a detailed form
\begin{equation}
	\frac{y^4}{\left( y^{4}(1+y^{2}\ell^{2})-\frac{2M}{\Xi^{2}}y^{2} + \frac{2Ma^{2}}{\Xi^{3}}\right) }\,\dot{y}^2 \,=\,\kappa^2\,\left( 1+ y^2 \ell^2 - \frac{2M}{y^2 \Xi^3} \right)- y^2\,K^2\,, 
\end{equation}
where
\begin{equation}\label{Kdef}
	K^2\,\equiv\, m_{\phi}^2\,\sin^2\Theta^{*} +  m_{\psi}^2\, \cos^2\Theta^{*}   \,.
\end{equation}
The above equation can be represented as in the following form
\begin{equation}
	\left( \frac{\dot{y}}{y^2}\right) ^2 \,=\, \left[ \kappa^2 \,\left(-\frac{2M}{\Xi^3}\,\frac{1}{y^4}+\frac{1}{y^2}+\ell^2\right) -K^2 \right] \,\left[ \frac{2Ma^2}{\Xi^3}\,\frac{1}{y^6}    -\frac{2M}{\Xi^2}\,\frac{1}{y^4}+\frac{1}{y^2}+\ell^2 \right],\,
\end{equation}
or, equivalently
\begin{equation}\label{turningPnew}
	\dot{y}^2 \,=\, \left(\kappa^2\left(1+\ell^2 y^2-\frac{2M}{\Xi^3 y^2} \right)-y^2  K^2\right)\left(1+\ell^2 y^2-\frac{2M}{\Xi^2 y^2} + \frac{2Ma^2}{\Xi^3 y^4} \right).\,
\end{equation}
The first multiplier in (\ref{turningPnew}) has the following roots
{\footnotesize
\bea\label{turnPFM}
y_{1} =  \frac{ 2|\kappa| M}{\sqrt{\kappa^2 \Xi ^3+\sqrt{\kappa^2 \Xi ^3 (\kappa^2 (8\ell^2 M+\Xi^3)-8 K^2 M)}}} ,\quad y_{2} = \frac{2|\kappa| \sqrt{M}}{\sqrt{\kappa ^2 \Xi ^3-\sqrt{\kappa ^2 \Xi ^3 (\kappa ^2 (8 \ell^2 M+\Xi ^3)-8 K^2 M)}}}, \,\\ \label{turnPFM2}
 y_{3}= \frac{-2|\kappa| \sqrt{M}}{\sqrt{\kappa ^2 \Xi^3-\sqrt{\kappa^2 \Xi^3 (\kappa ^2  (8 \ell^2 M+\Xi ^3)-8 K^2 M)}}}, \quad y_{4} =\frac{-2|\kappa|\sqrt{ M}}{\sqrt{\kappa ^2 \Xi ^3+\sqrt{\kappa ^2 \Xi ^3 (\kappa ^2
   (8 \ell^2 M+\Xi ^3)-8 K^2 M)}}}.\,.
\eea}
We are able to tune parameters  $M$, $a$,  $K$ and $\kappa$ in such a way that there will be two real positive roots $y_1$ and $y_2$ (\ref{turnPFM}).

     \begin{figure}[t!]
\centering
\includegraphics[width=7cm]{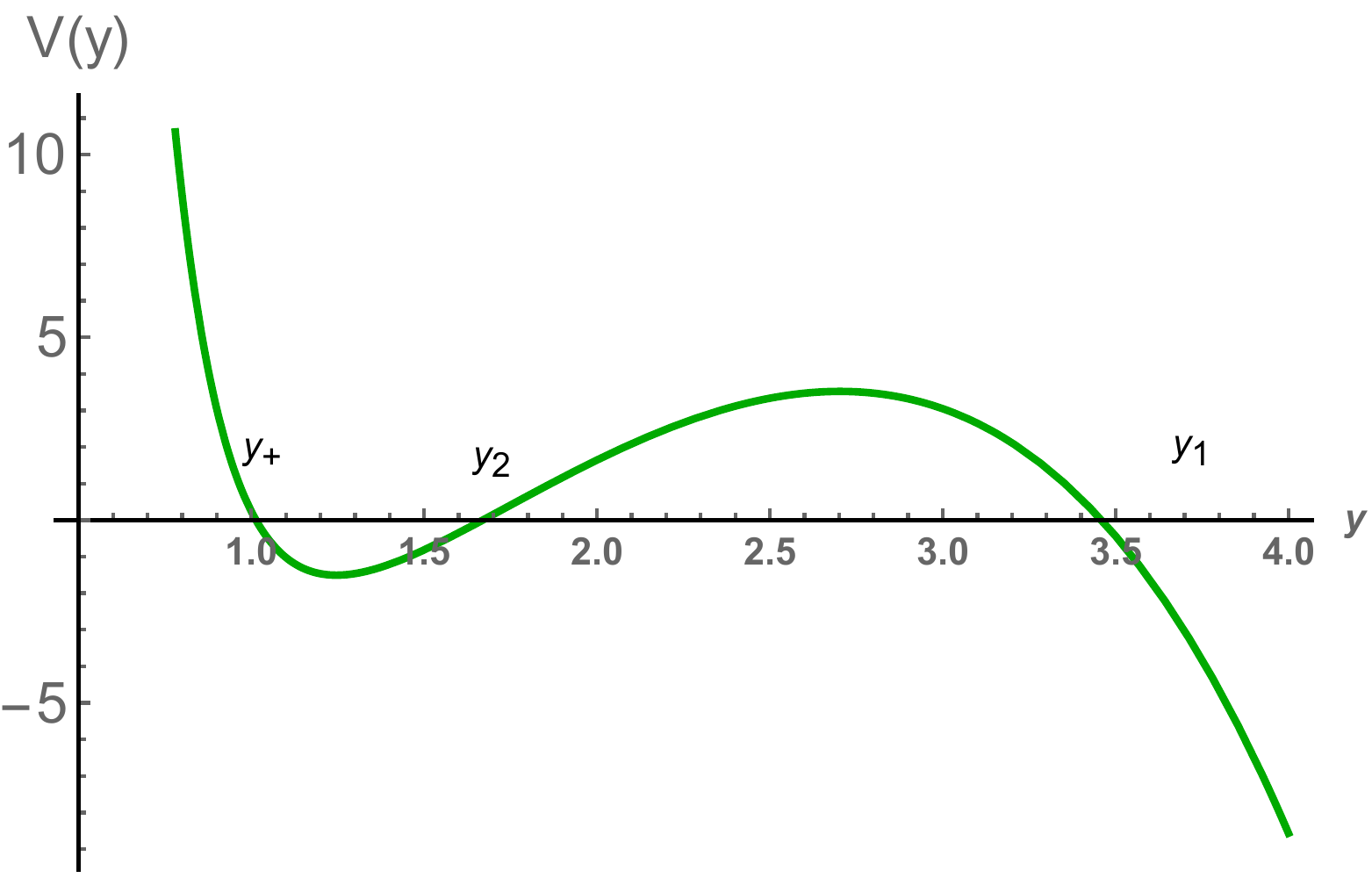} 
    \caption{The behaviour of the effective potential of the system (\ref{equationYtp}). The parameters are fixed as follows: $M=1$, $a=0.2$, $K^2=1.55$, $\kappa^2=1.5$, $y_{+}=1.01347$, $y_{1} =3.45935$, $y_{2} =1.66944$.} \label{fig:Vykappa}
\end{figure}

As for the second multiplier in (\ref{turningPnew}), it is nothing but the blackening function 
with a greater root, which is the horizon $y_{+}$ of the Kerr-$AdS_{5}$ black hole. The zeroes of this function are presented in the Appendix~(\ref{rootbf1})-(\ref{rootbf6}). It worth to be noted that four of them are complex.

Then, the equation \eqref{turningPnew} can be represented in the following  form

\begin{equation}\label{equationYtp}
		\dot{y}^2\,= \kappa^2 \left(\ell^2-\frac{ K^2}{\kappa^2}\right)\frac{\ell^2}{y^{6}}(y-y_{-})(y-y_{+})\displaystyle{\prod_{j = 1}^{4}(y- y_{j})(y-y_{j}^{*})},
\end{equation}
where $y_{-}$, $y_{+}$ are real zeros of blackening function, thus $y_{+}$ (\ref{rootbf1}) is the horizon, $y_{1}^{*}, y^{*}_{2},y^{*}_{3},y^{*}_{4}$ are complex zeros of the blackening function given by (\ref{rootbf3})-(\ref{rootbf6}), 
while $y_{1}$, $y_{2}$, $y_{3}$ and $y_{4}$ (\ref{turnPFM})-(\ref{turnPFM2}) are zeros of the first multiplier of eq.(\ref{turningPnew}). So we can always find appropriate conditions on the right-hand side of (\ref{equationYtp}) for the existence of a periodic solution
\be
y_{-}<0<y_{+}<y_2<y(\tau)< y_1,
\ee
where $y_1$ and $y_2$ are given by (\ref{turnPFM}).
Therefore, there exists a pulsating string configuration, expanding and contracting between the turning points $\,y_{1}$ and $\,y_{2}\,$. 
In Fig.~\ref{fig:Vykappa} we plot the potential of the effective mechanical system (\ref{equationYtp}).  One can see that it has three positive real zeroes $y_{+}$, $y_{1}$, $y_{2}$. 
The evolution of the pulsating string is defined in the region  $\,y_{1}<y(\tau)<\,y_{2}\,$. 

We also note that $\Phi$ and $\Psi$ are defined by the relations
\be
\Phi =m_{\phi}\sigma, \quad \Psi= m_{\psi}\sigma.
\ee


\subsection{Pulsating string configuration on the subspace $y=const$} \label{Sect32}

In order to obtain pulsating string solutions on the subspace $y=const$ we
consider the following string ansatz ($\,\kappa > 0\, $):
\begin{equation}\label{ansatz}
	T =\kappa \tau, \quad \Theta = \theta(\tau),\quad {\Phi = m_{\phi}\sigma +\phi(\tau),\quad \Psi = m_{\psi}\sigma + \psi(\tau)}, \quad y = const.
\end{equation}
Taking into account \eqref{ansatz} the string Lagrangian can be represented as
\bea\label{lagrangianAnsatz}
\mathcal{L}_{P} &\sim& - \left( \kappa^{2} G_{TT} + \dot{\theta}^{2} G_{\Theta\Theta} 
+\dot{\phi}^{2}G_{\Phi\Phi} +\dot{\psi}^{2} G_{\Psi\Psi}  +  2 \kappa(\dot{\phi} G_{T\Phi} + \dot{\psi} G_{T\Psi}) +2 \dot{\phi}\dot{\psi} G_{\Phi\Psi}\right)\nonumber\\ 
& +&  m_{\phi}^{2} G_{\Phi\Phi}  +m_\psi^{2} G_{\Psi\Psi} +2 m_\phi m_{\psi} G_{\Phi\Psi}.
\eea
Let us list the relevant equations of motion (EoM). Since the ansatz \eqref{ansatz} is linear in worldsheet time $\,\tau\,$, the EoMs become actually equations with respect to $\,\sigma\,$, while all constants $\,A_T , \,\, A_{\phi},\,\, A_{\psi}\,$ below are the integration constants. The EoMs for $\,\Phi,\,\,\Psi\,$ and $\,T\,$ read off as follows
\bea
\Phi:\quad &-&\frac{2M}{y^{2}\Xi^{3}}\kappa +\left(\frac{y^{2}}{a}+\frac{2aM}{y^{2}\Xi^{3}}\sin^{2}\theta\right) \dot{\phi}  + \frac{2Ma\cos^{2}\theta }{\Xi^{3}y^{2}}\dot{\psi} = \frac{A_{\phi}}{a\sin^{2}\theta},\\
\Psi:\quad &-&\frac{2M}{y^{2}\Xi^{3}}\kappa +\left(\frac{y^{2}}{a}+\frac{2aM}{y^{2}\Xi^{3}}\cos^{2}\theta\right) \dot{\psi}  + \frac{2Ma\sin^{2}\theta }{\Xi^{3}y^{2}}\dot{\phi} = \frac{A_{\psi}}{a\cos^{2}\theta},\\
T:\quad  && \frac{2M}{y^{2}\Xi^{3}}\kappa  -\frac{2aM}{y^{2}\Xi^{3}}(\dot{\phi}\sin^{2}\theta + \dot{\psi} \cos^{2}\Theta) = A_{T}+ (1 + y^{2}l^{2})\kappa.
\eea
Correspondingly, the equation for $\,y\,$ has the form
\bea\label{yeq2}
&& \kappa^{2}(2\ell^{2}y+ \frac{4M}{y^{3}\Xi^{3}}) - 2y \dot{\theta}^{2} - \frac{4\kappa}{y}\left(\frac{2Ma\sin^{2}\theta\dot{\phi}}{y^{2}\Xi^{3}} + \frac{2Ma\cos^{2}\theta\dot{\psi}}{y^{2}\Xi^{3}}\right)\nonumber\\
&+&2y \left(\sin^{2}\theta m^{2}_{\phi} + m^{2}_{\psi}\cos^{2}\theta\right) - 2y(\sin^{2}\theta\dot{\phi}^{2}+ \cos^{2}\theta\dot{\psi}^{2})\nonumber\\
&-& \frac{4Ma^{2}}{y^{3}\Xi^{3}}\Bigl(-\sin^{4}\theta\dot{\phi}^{2} - \cos^{4}\theta\dot{\psi}^{2} - 2\sin^2\theta\cos^2\theta\dot{\phi}\dot{\psi}\nonumber\\
&+&m^{2}_{\phi} \sin^{4}\theta + m^{2}_{\psi}\cos^{4}\theta+ 2m_{\phi}m_{\psi}\cos^{2}\theta\sin^{2}\theta\Bigr)=0.
\eea
The equation for $T$ gives the ratio
\be\label{pconst1}
\dot{\phi}\sin^{2}\theta + \dot{\psi} \cos^{2}\theta = \frac{y^{2}\Xi^{3}}{2aM}\left(\frac{2M}{y^{2}\Xi^{3}}\kappa - A_{T} - (1+y^{2}\ell^{2})\kappa\right).
\ee
Combining the equations for $\Phi\,$, $\Psi\,$ and $T$  and doing some algebra we obtain the relevant equations for $\,\phi\,$ and $\,\psi\,$
\bea
T +\Phi: &-& (1 + y^{2}l^{2})\kappa + \frac{y^{2}}{a}\dot{\phi} =  \frac{A_{\phi}}{a\sin^{2}\theta} + A_{T},\nonumber \\ 
\dot{\phi}&=&\frac{1}{y^{2}}\frac{A_{\phi}}{\sin^{2}\theta} +\frac{a}{y^{2}} \left(A_{T} + (1 + y^2\ell^{2})\kappa\right),\label{eqphi1}\\
T+\Psi:&-& (1 + y^{2}l^{2})\kappa + \frac{y^{2}}{a}\dot{\psi}  =  \frac{A_{\psi}}{a\sin^{2}\theta} + A_{T},\nonumber\\ 
\dot{\psi}&=&\frac{1}{y^{2}}\frac{A_{\psi}}{\cos^{2}\theta} + \frac{a}{y^{2}}\left(A_{T} + (1 + y^2\ell^{2})\kappa\right). \label{eqpsi1}
\eea
Putting the expressions \eqref{eqphi1} and \eqref{eqpsi1} into the equation for $T$ \eqref{pconst1} we obtain the relation between constants
\be\label{relation T}
\frac{2M}{y^{2}\Xi^{3}}\kappa -\left(A_{\phi} + A_{\psi}\right) \frac{2aM}{y^{4}\Xi^{3}} - (1  +  \frac{2a^{2}M}{y^{4}\Xi^{3}}) \left(A_{T} + (1 + y^2\ell^{2})\kappa\right) =0.
\ee
The equations for the functions  $\,\phi\,$ and $\,\psi\,$ can be rewritten in the form
\bea
\dot{\phi}&=&\frac{A_{\phi}}{y^{2}\sin^{2}\theta} + P - \frac{1}{y^{2}}(A_{\phi} + A_{\psi}),\label{eq phi}\\
\dot{\psi}&=&\frac{A_{\psi}}{y^{2}\cos^{2}\theta} + P - \frac{1}{y^{2}}(A_{\phi} + A_{\psi}),\label{eq psi}
\eea
where the constant $P$ is given by
\begin{equation}\label{P}
	P=\frac{1}{y^{2}}A_{\phi} +\frac{1}{y^{2}}A_{\psi} +\frac{a}{y^{2}} \left(A_{T} + (1 + y^2\ell^{2})\kappa\right).
\end{equation}
From here we find
\be\label{relation sin cos}
\dot{\phi}\sin^{2}\theta +\dot{\psi}\cos^{2}\theta= P, \quad A_{T}  + (1 + y^{2} \ell^{2})\kappa = -\frac{2M(a(A_{\phi} + A_{\psi}) - \kappa y^{2})}{2a^{2}M + y^{4}\Xi^{3} }.
\ee
Substituting (\ref{pconst1}) in (\ref{yeq2}) with $\frac{1}{y}$ factor we obtain the $y$-equation in the form
\bea
&& \frac{\kappa^{2}}{y}\left(\ell^{2}y+ \frac{2M}{y^{3}\Xi^{3}}\right) -  \dot{\theta}^{2} - \frac{2\kappa}{y^{2}}\left(\frac{2M\kappa}{y^{2}\Xi^{3}} - (A_{T} +(1+y^{2}\ell^{2})\kappa)\right)\nonumber\\
&+& \left(\sin^{2}\theta m^{2}_{\phi} + m^{2}_{\psi}\cos^{2}\theta\right) - (\sin^{2}\theta\dot{\phi}^{2}+ \cos^{2}\theta\dot{\psi}^{2})\nonumber\\
&-&\frac{2Ma^{2}}{y^{4}\Xi^{3}}\left(-(\sin^{2}\theta\dot{\phi} +\cos^{2}\theta\dot{\psi})^{2} + (m_{\phi}\sin^{2}\theta + m_{\psi}\cos^{2}\theta)^{2}\right) =0.\label{eq y}
\eea
The	first Virasoro constraint \eqref{Vir1} explicitly reads
\bea\label{Vir1theta}
&& - \frac{\kappa^{2}}{y^{2}}\left(1 + y^{2}l^{2}-\frac{2M}{y^{2}\Xi^{3}}\right)- \frac{4aM}{y^{4}\Xi^{3}}\kappa(\sin^{2}\theta\dot{\phi} + \cos^{2}\theta\dot{\psi}) + \dot{\theta}^{2}\nonumber\\
&&+ (\sin^{2}\theta \dot{\phi}^{2} +\cos^{2}\theta\dot{\psi}^{2}) + (\sin^{2}\theta m_{\phi}^{2}+\cos^{2}\theta m^{2}_{\psi}) \nonumber\\
&&+ 2\frac{Ma^{2}}{\Xi^{3}y^{4}}\left((\sin^{2}\theta\dot{\phi} + \cos^{2}\theta\dot{\psi})^{2} + (\sin^{2}\theta m_{\phi} + \cos^{2}\theta m_{\psi} )^{2} \right) = 0.
\eea
Remembering the equation for $\,y\,$ \eqref{eq y} and expression for the constant $\,P\,$ \eqref{P} and summing equations \eqref{eq y} and \eqref{Vir1theta} we obtain the constraint
\bea\label{y+vir1}
-\frac{\kappa^{2}}{y^{2}} + 4\frac{\kappa^2M}{y^{4}\Xi^{3}} - 8\frac{aM}{y^{4}\Xi^{3}}\kappa P +2 \left(m^{2}_{\phi}\sin^{2}\theta +m^{2}_{\psi}\cos^{2}\theta \right)  + \frac{4Ma^{2}}{y^{4}\Xi^{3}}P^{2} =0.
\eea
Since, $\,\theta\,$ is essentially time dependent (\ref{ansatz}), we are forward to impose the condition 
\begin{equation}\label{condition m}
	m^{2}_{\phi}\,=\,m^{2}_{\psi}\,\equiv\,m^2.
\end{equation}	
With this choice the second Virasoro constraint \eqref{Vir2} give us the relation
\begin{equation}\label{Vir2 m}
	\textrm{Vir2}:\qquad-\frac{2M}{y^{2}\Xi^{3}}\kappa +(A_{\phi} + A_{\psi})(\frac{1}{a}+ \frac{2aM}{y^{4}\Xi^{3}}) + (1  +  \frac{2a^{2}M}{y^{4}\Xi^{3}}) \left(A_{T} + (1 + y^2\ell^{2})\kappa\right) =0.
\end{equation}
As a result, we find
\be\label{condition A}
\frac{A_{\phi}+A_{\psi}}{a} =0,  \quad A_{\phi}= -A_{\psi}\,\equiv\,A\,.
\ee
Thus, the constant $P$  \eqref{P} can be rewritten as
\be\label{conditionP}
P = \frac{a}{y^{2}}(A_{T} + (1+y^2\ell^2)\kappa).
\ee
Consequently,  by virtue of the above relations between the constants and expression for the second Virasoro constraint \eqref{Vir2 m} we can fix $\,A_T\,$ 
\be\label{A_T fix}
A_T\,=\,\frac{2M\kappa\,y^2}{2a^{2}M +y^{4} \Xi^{3}}-(1 + y^2\ell^{2})\kappa\,, \qquad P = \frac{2aM\kappa}{2a^{2}M +y^{4} \Xi^{3}}.
\ee
Having $m^{2}_{\phi}=m^2_{\psi} =m^2$ and with the assumptions (\ref{condition A}), the equation for the first Virasoro constraint \eqref{Vir1theta} takes the form
\be
\dot{\theta}^{2}- \frac{\kappa^{2}}{y^{2}}(1 + y^{2}l^{2}-\frac{2M}{y^{2}\Xi^{3}})-\frac{4aM}{y^{4}\Xi^{3}}\kappa P + (2\frac{Ma^{2}}{\Xi^{3}y^{4}}+1)(P^{2} + m^{2}) 
+ \frac{A^{2}}{y^4}\frac{1}{\sin^{2}\theta\cos^{2}\theta} = 0.
\ee
Finally, plugging the expression for $P$ \eqref{conditionP} we obtain the following differential equation for $\,\theta\,$
\be\label{thetaeqc}
\dot{\theta}^{2}+ \frac{A^{2}}{y^4}\frac{1}{\sin^{2}\theta\cos^{2}\theta} + \Upsilon  =0, 
\ee
where
\be
\Upsilon = - \frac{\kappa^{2}}{y^{2}}(1 + y^{2}l^{2}-\frac{2M}{y^{2}\Xi^{3}})+ (2\frac{Ma^{2}}{\Xi^{3}y^{4}}+1)m^2-\frac{4a^2M^2\kappa^2}{y^{4}\Xi^{3}(2Ma^{2}+y^{4}\Xi^{3})}.
\ee
At this point, it is convenient to introduce a new variable $u$ 
\be\label{ucoordpuls}
u(\tau) \equiv \sin^{2}\theta(\tau) >0, \quad 0<u< 1.
\ee
Then we have
\be\label{equUV}
\frac{\dot{u}^{2}}{2} + 2|\Upsilon|u^{2} - 2|\Upsilon|u+\frac{2A^{2}}{y^{4}}= 0,\qquad \Upsilon <0 .
\ee
The equation (\ref{equUV})  can be written in the form
\be
\frac{\dot{u}^{2}}{2} + V(u) = 0, \quad V(u) =  2|\Upsilon|u^{2} - 2|\Upsilon|u+\frac{2A^{2}}{y^{4}}.
\ee
If $\frac{4A^{2}}{|\Upsilon|^2\,y^{4}}\,<1\,$ the potential $V(u)$ has two turning points $\,	u_{1,2} = \frac{1}{2}\mp \frac{\sqrt{1-\frac{4A^{2}}{|\Upsilon|^2\,y^{4}}}}{2}\,$ symmetrical about the point of minimum of the potential $\,u=\dfrac{1}{2}\,$ and $\,V(u)\leq 0\,$, for $\,0<u_1 \leq\,u(\tau)\,\leq\,u_2<1\,$.
Therefore, the above equation
\be 
\dot{u}^{2} = - 4|\Upsilon|u^{2} + 4|\Upsilon|  u - \frac{4A^{2}}{y^{4}}\equiv\,-2V(u) \geq 0
\ee
has a periodic solution between turning points $\,u_{1,2}\,$.  The periodic solution of the above equation is
\be\label{puls solution}
u(\tau) = \frac{1}{2} + \sqrt{\frac{1}{4} - \frac{A^{2}}{|\Upsilon|y^{4}}}\sin(2\sqrt{|\Upsilon}|\tau).
\ee
It is a pulsating string solution moving between the turning points.
Correspondingly, the dynamics on the angular variables $\Phi$ and $\Psi$ is defined by \eqref{eq phi}-\eqref{eq psi},which with the constraint \eqref{condition A} and \eqref{ucoordpuls} take the form
\bea
\dot{\phi}&=&\frac{A}{y^{2}u(\tau)} + \frac{2aM\kappa}{2a^{2}M +y^{4} \Xi^{3}},\label{eq phi2}\\
\dot{\psi}&=&-\frac{A}{y^{2}(1-u(\tau))} +\frac{2aM\kappa}{2a^{2}M +y^{4} \Xi^{3}}. \label{eq psi2}
\eea

%


\section{Energy spectrum}\label{Sect4}
In this section we will semi-classically quantize the pulsating string configuration in 5d Kerr-AdS geometry. The approach has been presented for the first time in \cite{min}. We will calculate the corresponding energy spectra.
First, we will relate a study of the string energy spectra to the Bohr-Sommerfeld problem. After this, we discuss  the large energies (large quantum numbers) to  find the first correction to the energy. Note,  the  energy spectra of the circular string in the AdS background are related to the anomalous dimensions of the CFT operators according to the AdS/CFT dictionary.  However, for the case of the Kerr-AdS spacetime we cannot carry out this connection, but we are able to relate the energy spectra to the dispersion relations.

      \subsection{Bohr-Sommerfeld analysis in 5d Kerr-AdS}\label{Sect41}
      
We consider a circular closed string in the  5d  Kerr-AdS black hole (\ref{GLKA}). In this analysis the string dynamics is governed by the Nambu-Goto action  reads as
\be\label{SNGg}
S_{NG} = - \frac{1}{2 \pi \alpha'} \int  d \tau d\sigma \sqrt{|h|},
\ee
where the induced metric on the worldsheet 
\be\label{indm-h} 
h_{\alpha\beta} = G_{MN}\partial_{\alpha}X^{M}\partial_{\beta}X^{N},\quad \textrm{with}\quad X^{M} = (T,\Theta,\Phi, \Psi,y)\,.\ee
For the embedding we choose the  ansatz
\be\label{BSansatz}
	T=\kappa\,\tau,\qquad y=y(\tau),\qquad \Theta=\Theta^*\,=\,const,\qquad \Phi=\,m_{\phi}\sigma,\qquad \Psi=\,m_{\psi} \sigma\,.
\ee

The components of the induced metric (\ref{indm-h}) take the form
\bea\label{eq:indm1}
h_{\tau\tau}&=& G_{TT}(\partial_{\tau}T)^2 + G_{yy}(\partial_{\tau}y)^2  =\kappa^2\Bigl(-1-y^2\ell^2 + \frac{2M}{y^2 \Xi^3 }\Bigr) + \frac{\dot{y}_{\tau}^2}{f(y)},\\
h_{\sigma\sigma}&=& G_{\Phi \Phi} (\partial_{\sigma}\Phi)^2 + 2G_{\Phi \Psi} \partial_{\sigma} \Phi \partial_{\sigma} \Psi + G_{\Psi \Psi} (\partial_{\sigma} \Psi)^2  \nonumber\\
&=&y^{2}(\sin^{2}\Theta^{*} m^2_{\phi}+\cos^2\Theta^{*} m_{\psi}^2) + \frac{2Ma^2}{\Xi^{3}y^2}\left(\sin^2\Theta^*{} m_{\phi} + \cos^{2}\Theta^{*} m_{\psi}\right)^2,\\ \label{eq:indm3}
 h_{\sigma\tau}&=& 2G_{T \Phi}\partial_{\tau}T \partial_{\sigma} \Phi + 2G_{T \Psi} \partial_{\tau} T \partial_{\sigma} \Psi = -\frac{2\kappa Ma}{\Xi^3 y^2}\Bigr(\sin^2 \Theta^{*} m_{\phi} + \cos^2 \Theta^{*} m_{\psi} \Bigr).
\eea
By virtue of (\ref{eq:indm1})-(\ref{eq:indm3}) the  NG action (\ref{SNGg}) is written down as
{\footnotesize
\be
\label{NGyT}
S_{NG} = - \frac{1}{2\pi \alpha'} \int d\tau d\sigma \sqrt{\Bigr(\kappa^2\Bigl(1+ y^2\ell^2 - \frac{2M}{y^2 \Xi^3 }\Bigr) - \frac{\dot{y}_{\tau}^2}{f(y)} \Bigr)\cdot \Bigr( y^2 \tilde{\mathcal{M}}+ \frac{2Ma^2}{\Xi^3y^2} \mathcal{M}^2\Bigr) + \Bigr(\frac{2\kappa Ma}{\Xi^3 y^2}\mathcal{M}\Bigr)^2},
\ee}
where
\be\label{blackFfunc}
f(y) = 1+\frac{2Ma^2}{\Xi^3y^4} - \frac{2M}{\Xi^2y^2} +\ell^2y^2
\ee
and
\be
\mathcal{M} = m_{\phi}\sin^{2}\Theta^{*}+ m_{\psi}\cos^{2}\Theta^{*},\quad \tilde{\mathcal{M}} = m^{2}_{\phi}\sin^{2}\Theta^{*}+ m^2_{\psi}\cos^{2}\Theta^{*}. 
\ee
Doing some algebra, we can be represent the action (\ref{NGyT}) in the simplified  form
{\footnotesize
\be
S_{NG} = - \frac{1}{2\pi \alpha'} \frac{\kappa}{\Xi^{3/2}} \int d\tau  \sqrt{ (1+\ell^2 y^2)\left(y^2 \Xi^3 \tilde{\mathcal{M}} +\frac{2 M\mathcal{M}^2 a^2}{y^2} \right) -2 M\tilde{\mathcal{M}} -\frac{\dot{y}^2_{\tau}}{\kappa^2 f(y)}\left(y^2 \Xi^3 \tilde{\mathcal{M}} +\frac{2 M\mathcal{M}^2 a^2}{y^2}\right)}.
\ee}
The canonical momentum corresponding to (\ref{NGyT}) is
\be\label{canM}
\Pi = -\frac{\dot{y }(2 a^2 M \mathcal{M}^2+\tilde{\mathcal{M}} \Xi ^3 y^4)}{\kappa  \Xi ^{3/2} y f(y)
   \sqrt{(1+\ell^2 y^2) (2 a^2 M \mathcal{M}^2+\tilde{\mathcal{M}} \Xi ^3 y^4)-2 M\tilde{\mathcal{M}}y^2-\frac{\dot{y}^2 \left(2 a^2 M \mathcal{M}^2+\tilde{\mathcal{M}} \Xi ^3 y^4\right)}{\kappa ^2 f(y)}}}.
\ee
and the first integral related to (\ref{NGyT}) is given by
\be \label{Ham}
\mathcal{H} = -\frac{\kappa  \left( (\tilde{\mathcal{M}} \Xi ^3 y^4 +2 a^2 M \mathcal{M}^2 )(1+\ell^2y^2)-2 M \tilde{\mathcal{M}} y^2\right)}{\Xi ^{3/2} y \sqrt{(1 + \ell^2 y^2) (2 a^2 M \mathcal{M}^2+\tilde{\mathcal{M}} \Xi^3 y^4)-2 M \tilde{\mathcal{M}}y^2-\frac{\dot{y}^2 \left(2 a^2 M   \mathcal{M}^2+\tilde{\mathcal{M}} \Xi^3 y^4\right)}{f(y) \kappa^2 }}}.
\ee

 It is convenient to pass to a new variable\footnote{{\it Comment on the change of the variable $\xi = \int \mathbf{f}(y)dy$ $\Rightarrow$ $\xi= \mathbf{F}(y)$  $\Rightarrow$  $\xi = \mathbf{F}(y(\tau))$  $\Rightarrow$ $\dot{\xi} = \frac{\partial \mathbf{F}}{\partial y}\dot{y}$ $\Rightarrow$ $\xi = \mathbf{f}(y)\dot{y}$}.}
\be\label{xivar}
\xi = \int \frac{dy}{\sqrt{\kappa f(y)}}\sqrt{\frac{(y^2 \Xi^3 \tilde{\mathcal{M}} +\frac{2 M\mathcal{M}^2 a^2}{y^2})}{(1+\ell^2 y^2)(y^2 \Xi^3 \tilde{\mathcal{M}} +\frac{2 M\mathcal{M}^2 a^2}{y^2} ) -2 M\tilde{\mathcal{M}}}}.
\ee

In terms of the $\xi$-variable (\ref{xivar}) the string Lagrangian (\ref{NGyT}) significantly simplifies
\be
L_{s} \sim \frac{g(\xi)}{\Xi^{3/2}}\sqrt{1 - \dot{\xi}^{2}}, 
\ee
where the function $g(\xi)$ is defined as
\be
g(\xi) = \sqrt{(1+\ell^2 y^2)\Bigl(y^2 \Xi^3 \tilde{\mathcal{M}} +\frac{2 M\mathcal{M}^2 a^2}{y^2} \Bigr) -2 M\tilde{\mathcal{M}}}.
\ee
The Hamiltonian and the canonical momenta take the form, correspondingly
\be
\mathcal{H}_{\xi} = -\frac{g(\xi)}{\Xi ^{3/2} \sqrt{1-\dot{\xi}^2}},\quad \Pi_{\xi} = -\frac{g(\xi) \dot{\xi }}{\Xi ^{3/2} \sqrt{1-\dot{\xi}^2}}
\ee
Then the system can be described as 
\be
\mathcal{H}_{\xi}  = \sqrt{\Pi^2_{\xi} + V_{\xi}}, 
\ee 
where the potential is given by
\be\label{Potxi}
V_{\xi} = \frac{g(\xi)^2}{\Xi^{3}}.
\ee

\begin{figure}[t!]
\centering
\includegraphics[width=6cm]{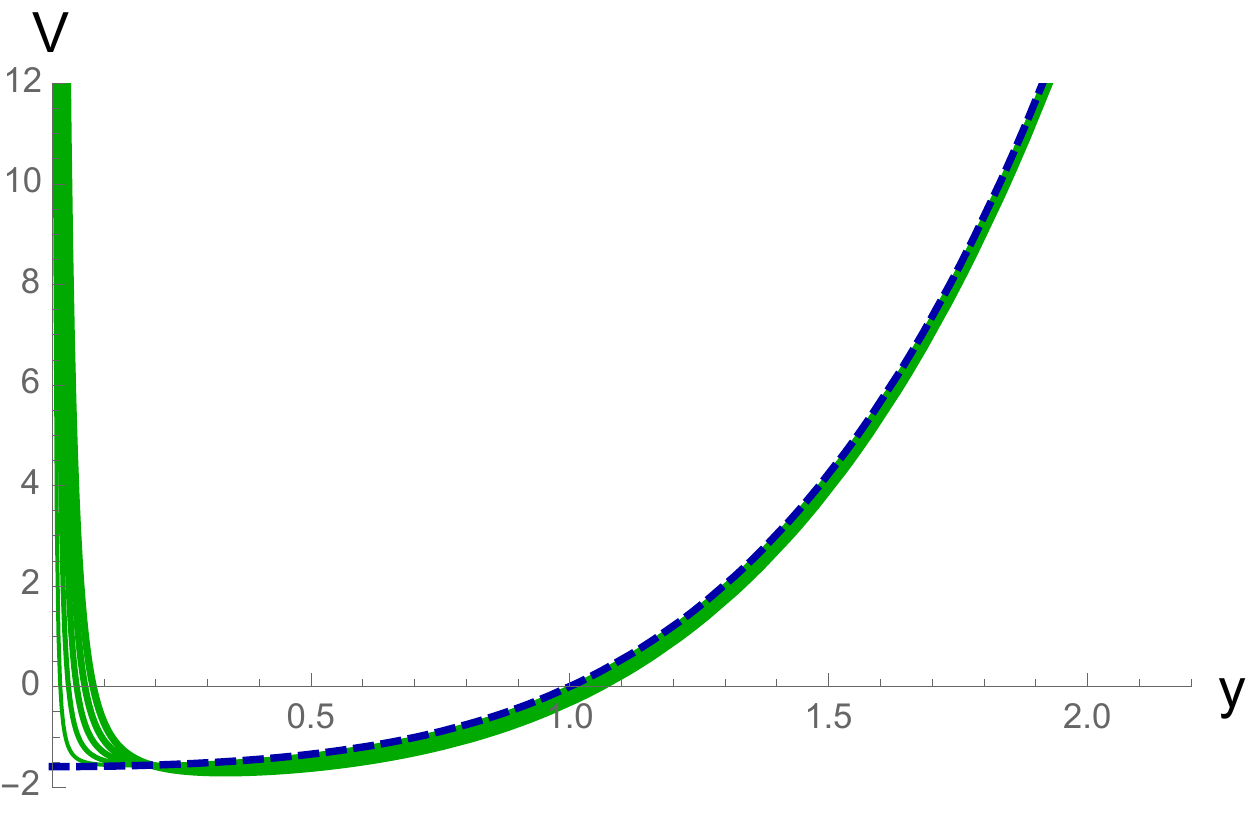}\,\includegraphics[width=1.5cm]{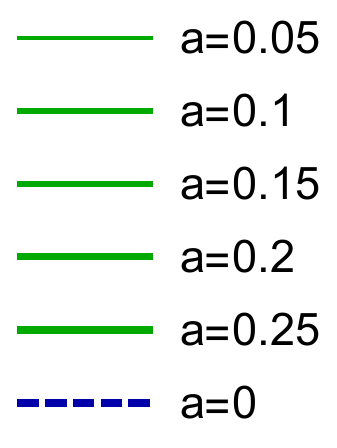}{\bf a}$\,$
\includegraphics[width=6cm]{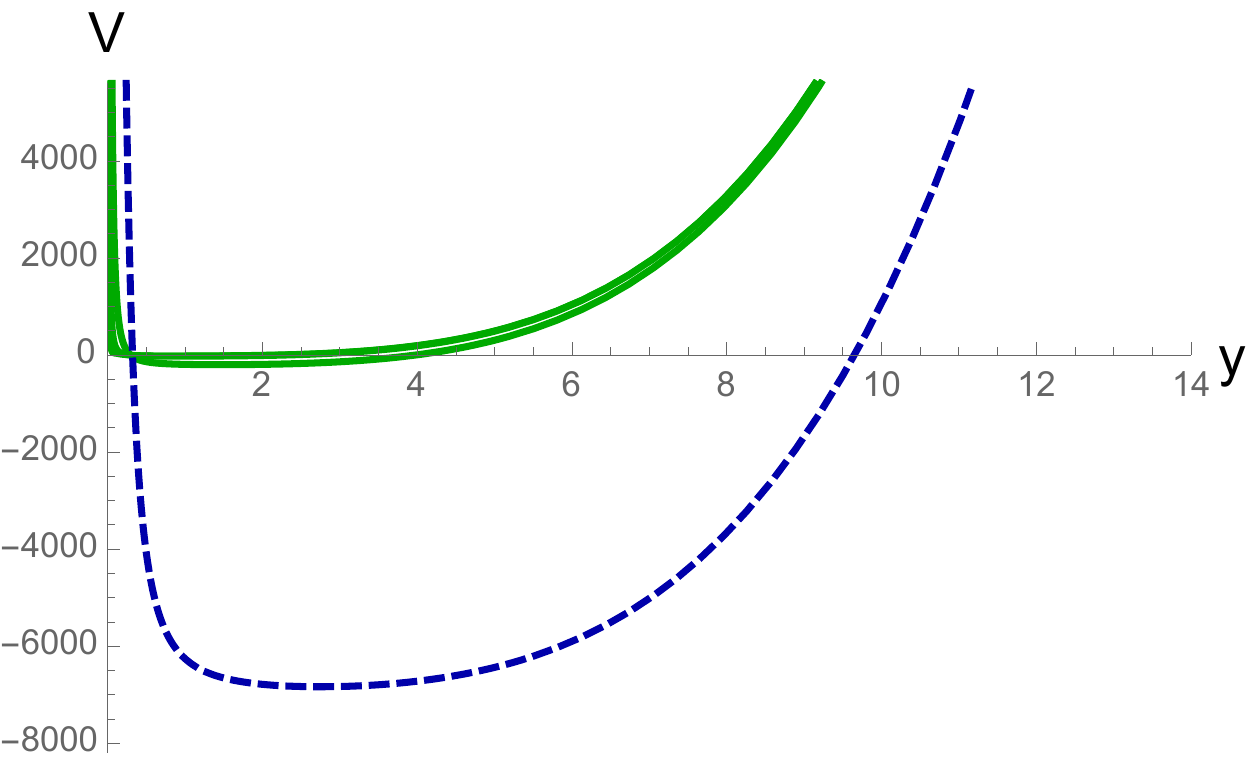}\,\includegraphics[width=1.6cm]{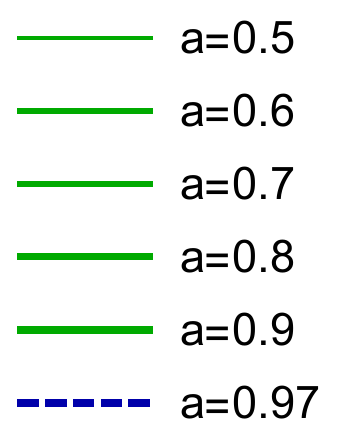}{\bf b}$\,$\\
$\,\,\,$\\
\includegraphics[width=6cm]{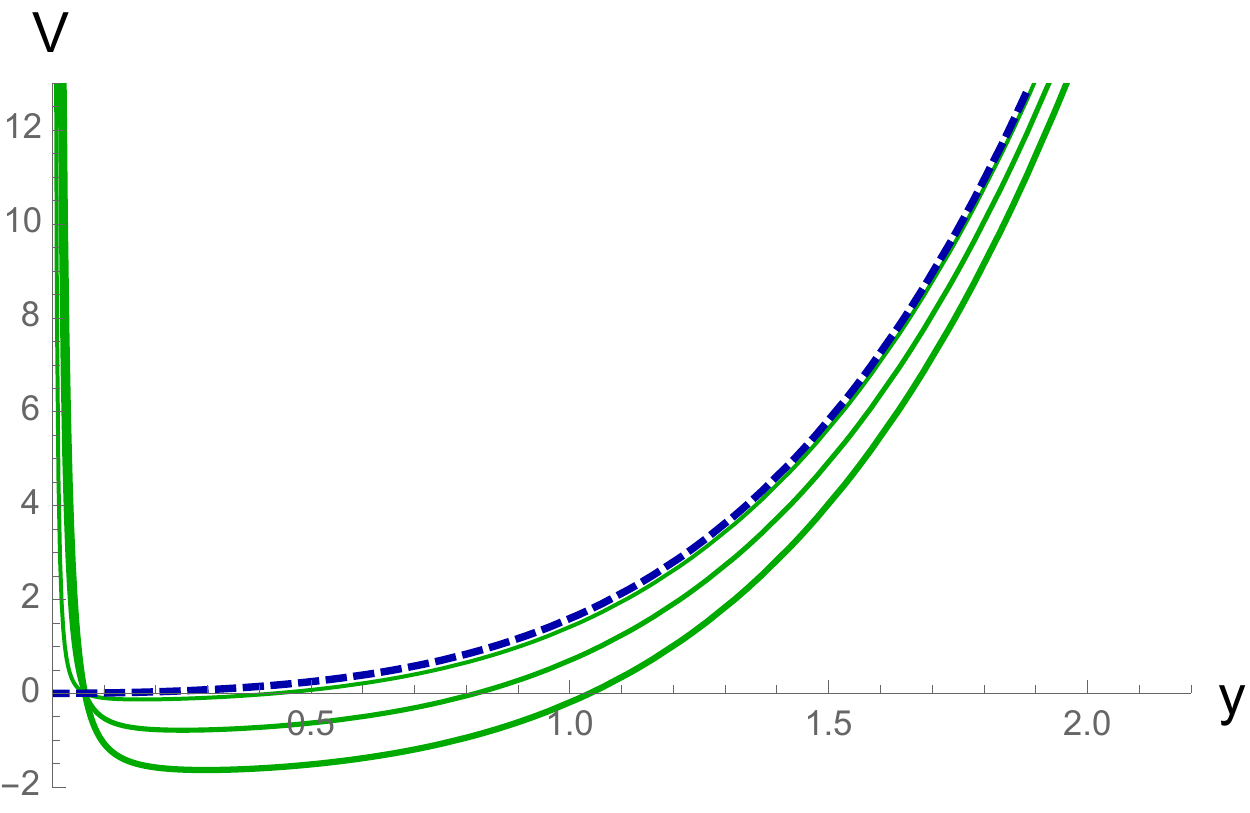}\,\includegraphics[width=2cm]{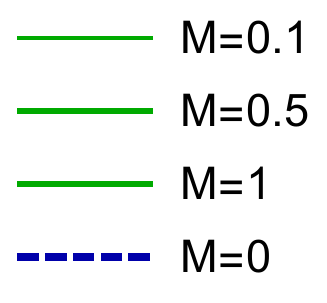}{\bf c}$\,$
    \caption{ The dependence of the potential $V$ on the radial coordinate $y$: {\bf a), b)} we keep fixed the mass $M$ and vary the rotational parameter $a$; {\bf c)} keep fixed the rotational parameter $a$ and change the mass $M$. We also keep $m_{\phi}=1$, $m_{\psi}=-2$ and $\Theta^{*} = \pi/4$.}
 \label{fig:Vya}
\end{figure}

The behaviour of the potential $V$ (\ref{Potxi}) is presented in Fig.~\ref{fig:Vya}.  In Fig.~\ref{fig:Vya} {\bf a)} and {b)} we show the potential for the fixed mass of the black hole varying  the rotational parameter, the case {\bf a)} corresponds to a small rotational parameter, while {b)} is for $a\in (0.5; 0.97)$. We recall that $a=1$ is a critical value of the rotational parameter for $\ell = 1$. In Fig.~\ref{fig:Vya} {\bf b)} we observe a jump of the potential for $a=0.97$.

The horizon of the  black hole $y_{+}$  and the boundary of the background can be  considered as potential wells.  So we can perform the
Bohr-Sommerfeld analysis for the quantization.
Then we have the following condition
\be\label{BSconstr}
(n + \frac{1}{2})\pi = \int^{\xi_{2}}_{\xi_{1}}d\xi \sqrt{E^2 - \frac{g(\xi)^2}{\alpha'^2\Xi^3}},
\ee
with the turning point $\xi_{1,2}$. 

In terms of the holographic radial coordinate $y$ (\ref{BSconstr}) takes the form
\bea\label{EIntnew}
&&\left(n+\frac{1}{2}\right)\pi = \int^{y_{1}}_{y_{+}}\frac{Edy}{\sqrt{\kappa f(y)}}\sqrt{\frac{(y^2 \Xi^3 \tilde{\mathcal{M}} +\frac{2 M\mathcal{M}^2 a^2}{y^2})}{(1+\ell^2 y^2)(y^2 \Xi^3 \tilde{\mathcal{M}} +\frac{2 M\mathcal{M}^2 a^2}{y^2} ) -2 M\tilde{\mathcal{M}}}} \times \nonumber\\
&&\sqrt{1-  \frac{(1+\ell^2 y^2)\Bigl(y^2 \Xi^3 \tilde{\mathcal{M}} +\frac{2 M\mathcal{M}^2 a^2}{y^2} \Bigr) -2 M\tilde{\mathcal{M}}}{\alpha'^2 \Xi^3 E^2}} \nonumber \\ \label{EIntnew3}
 &&= E\int^{y_{1}}_{y_{+}}dy\sqrt{Q(y)}\left(\frac{1}{\sqrt{\kappa f(y)}\sqrt{P(y)}} -\frac{1-\sqrt{1-\frac{1}{B^2\Xi^{3}\tilde{\mathcal{M}}}P(y)}}{\sqrt{\kappa f(y)}\sqrt{P(y)}}\right),
\eea
where we define $B = \frac{E\alpha'}{\sqrt{\tilde{\mathcal{M}}}}$ and 
\bea
P(y) &= &(1+\ell^2 y^2)\left(y^2 \Xi^3 \tilde{\mathcal{M}} +\frac{2 M\mathcal{M}^2 a^2}{y^2} \right) -2 M\tilde{\mathcal{M}},\\
 Q(y) &= &y^2 \Xi^3 \tilde{\mathcal{M}} +\frac{2 M\mathcal{M}^2 a^2}{y^2}.
 \eea

We are not able to calculate eq.~(\ref{EIntnew3}) exactly. However, supposing that the rotational parameter $a$ is small  we can expand in series (\ref{EIntnew3}), then we get
{\footnotesize
\bea\label{FTI}
&&\left(n+\frac{1}{2}\right)\pi= \frac{E}{\sqrt{\kappa}}\int^{\tilde{y}_{1}}_{y_{H}}dy\left(\frac{ 1}{1-\frac{2M}{y^2} + \ell^2 y^2} - \frac{1- \sqrt{1-\frac{y^2}{B^{2}}(1+\ell^2y^2 -\frac{2M}{y^2})}}{1-\frac{2M}{y^2} + \ell^2 y^2}\right) +\\ \label{STI}
 &&\frac{2 \tilde{\mathcal{M}} y^2 ( (5 \ell^2 y^2-1)- \frac{y^2}{B^2}(2 \ell^2 y^2-1) (1+\ell^2 y^2-\frac{2M}{y^2}))-3 \mathcal{M}^2 (2  M+\frac{y^4}{B^2}(1+\ell^2 y^2-\frac{2 M}{y^2})^2)}{2  \sqrt{\kappa } \tilde{\mathcal{M}}
  y^6 (\ell^2 y^2-\frac{2 M}{y^2} +1)^2 \sqrt{1-\frac{y^2}{B^2} (\ell^2 y^2+1-\frac{2 M}{y^2})}}a^2 M.
 \nonumber
 \eea}
 In eq. (\ref{FTI}) we observe the contribution
 \be\label{AdSSchwF}
h(y) = 1-\frac{2M}{y^2} + \ell^2 y^2,
\ee
which is nothing but the  blackenning function of the AdS-Schwarzschild black hole, i.e. the black hole without rotation.  The horizon $y_{H}$ for the the AdS-Schwarzschild black hole is defined as  a root of the equation  $h(y)= 0$, i.e.
\be\label{AdSSchwY}
 y_{H} = \frac{1}{\ell} \sqrt{\frac{1}{2}\Bigr(1 + \sqrt{1+8M\ell^2}\Bigr)}.
\ee
It is worth to note that the quantities $h(y)$ (\ref{AdSSchwF}) and $y_{H}$ (\ref{AdSSchwY})  related to the non-rotating  AdS black hole appear due to expanding around a small rotational parameter $a$.  

Integrating the first term in the integral \eqref{FTI} we get
\bea\label{ESpBSnew}
\frac{E}{\sqrt{\kappa}}\int^{\tilde{y}_{1}}_{y_{H}}dy\frac{ 1}{1-\frac{2M}{y^2} + \ell^2 y^2} = -\frac{E}{\sqrt{\kappa}} \frac{\sqrt{\sqrt{1+ 8\ell^2 M}-1}\tanh^{-1}\left(\frac{\sqrt{2}\ell y}{\sqrt{\sqrt{1+8\ell^2 M}-1}}\right)}{\ell\sqrt{2}\sqrt{1+8\ell^2 M}}\Big|^{\tilde{y}_{1}}_{y_{H}}\\
+  \frac{E}{\sqrt{\kappa}}\frac{2\sqrt{M} \tan^{-1}\left(\frac{\sqrt{2}\ell y}{\sqrt{\sqrt{1+8\ell^2 M}+1}}\right)}{\sqrt{1+8\ell^2 M}\sqrt{\sqrt{1+8\ell^2 M}-1}}\Big|^{\tilde{y}_{1}}_{y_{H}}.\nonumber
\eea
In the case of large energies $B\cdot \ell \gg 1$ eq.(\ref{ESpBSnew}) can be approximated as follows
\be\label{ESpBSnewapp}
\int^{\tilde{y}_{1}}_{y_{H}}dy\frac{ 1}{1-\frac{2M}{y^2} + \ell^2 y^2} \approx \frac{\pi}{2}\sqrt{\frac{1}{\sqrt{2 M}\ell^3}} -\sqrt{\frac{1}{B\ell^{3}}}.
\ee
The second integral \eqref{STI} can be calculated easily in terms of a new variable $y = r\sqrt{B/\ell}$
\bea\label{ESpBSnewappn}
\int^{y_{1}}_{y_{H}} dy\frac{1- \sqrt{1- \frac{y^2}{B^2}\left(1+ \ell^2 y^2 - \frac{2M}{y^2}\right)}}{1+ y^2 \ell^2 - \frac{2M}{y^2}} &\rightarrow& \sqrt{\frac{1}{\ell^3B}}\int^{1}_{0}\frac{dr}{r^2}(1 -\sqrt{1-r^4}) \nonumber\\\label{EnSpectBSapp2}
&=&\sqrt{\frac{1}{\ell^3 B}}\left(-1+\frac{(2\pi)^{3/2}}{\Gamma(\frac{1}{4})^2}\right).
\eea

Applying the same procedure to the third integral (\ref{STI}), in the limit of large $B$ we get
{\footnotesize
 \bea
 && \int^{\tilde{y}}_{y_{+}} dy \frac{2 \tilde{\mathcal{M}} y^2 ( (5 \ell^2 y^2-1)- \frac{y^2}{B^2}(2 \ell^2 y^2-1) (1+\ell^2 y^2-\frac{2M}{y^2}))-3 \mathcal{M}^2 (2  M+\frac{y^4}{B^2}(1+\ell^2 y^2-\frac{2 M}{y^2})^2)}{2  \tilde{\mathcal{M}}
  y^6 (\ell^2 y^2-\frac{2 M}{y^2} +1)^2 \sqrt{1-\frac{y^2}{B^2} (\ell^2 y^2+1-\frac{2 M}{y^2})}}a^2 M
 \nonumber\\
 & =& -\frac{\sqrt{ \ell}}{B^{5/2}}  Ma^2\int^{1}_{0}dr\left(\frac{2}{ r^2 \sqrt{1- r^4}} -\frac{5}{r^6 \sqrt{1- r^4}}+\frac{3 \mathcal{M}^2 }{2 \tilde{\mathcal{M}} r^2 \sqrt{1- r^4}}\right) \nonumber\\ \label{EnSpectBSappa}
 &=& -\frac{\sqrt{ \ell}}{B^{5/2}}\left(\frac{\sqrt{2} \pi ^{3/2}}{\Gamma (\frac{1}{4})^2}+\frac{3 \sqrt{\pi } \mathcal{M}^2 \Gamma (\frac{3}{4})}{2 \tilde{\mathcal{M}} \Gamma(\frac{1}{4})}\right)  Ma^2.
 \eea
}

 Combining together (\ref{ESpBSnewapp}),(\ref{EnSpectBSapp2}) and (\ref{EnSpectBSappa}) we get for Bohr-Sommerfeld quantisation condition 
 {\small
\be\label{EnwBSnew}
(n + \frac{1}{2})\pi \approx \frac{1}{\ell^{3/2}\sqrt{\kappa}} \left(\frac{\pi}{2}\frac{E}{(\sqrt{2M})^{1/2}} - \frac{(2\pi)^{3/2}}{\Gamma(\frac{1}{4})^2} \frac{(\tilde{\mathcal{M}} E)^{1/2}}{ (\alpha')^{1/2}} - \frac{\ell^2\sqrt{\pi } \tilde{\mathcal{M}}^{5/4}}{\alpha ^{5/2} E^{3/2}}\left(\frac{\sqrt{2} \pi}{\Gamma (\frac{1}{4})^2}+\frac{3  \mathcal{M}^2 \Gamma (\frac{3}{4})}{2 \tilde{\mathcal{M}} \Gamma(\frac{1}{4})}\right) 
Ma^{2}\right).
\ee}
In (\ref{EnwBSnew}) we can recognize the result for the energy from \cite{Alishahiha:2002fi} if one sends $a$ to $0$.

\subsection{WKB approximation}\label{Sect42}
\subsubsection{Derivation of the Hamiltonian}

As in the previous section the string dynamics is governed by the Nambu-Goto action (\ref{SNGg}) with the induced metric (\ref{indm-h}).
The first ingredient towards finding the spectrum is to make a pullback of the line element of the metric of the 5d Kerr-AdS background \eqref{GLKA} to the subspace, where string dynamics takes place. 
In this section one can consider more general ansatz than \eqref{ansatz} and \eqref{BSansatz}, such as
\begin{align}\label{new ansatz}
	&\Theta\equiv\,\xi_1=\xi_1(\tau)\,,\, y\equiv\,\xi_2=\xi_2(\tau)\,, \\
	&T\equiv\,X_0=x_0(\tau)+m_0\sigma,\qquad  x_0(\tau)=\kappa\tau, \qquad m_0=0\,,\\
	&\Phi\equiv\,X_1=m_1\sigma+x_1(\tau)\,, \quad \Psi\equiv\,X_2=m_2\sigma+x_2(\tau)\,.
\end{align} 
For convenience, we use the following notations of the Kerr-AdS metric \eqref{GLKA} 
\be\label{pulsMetric}
	ds^2 
	=\sum\limits_{i,j=1}^2 g_{ij}d\xi_i d\xi_j\,
	+\,\sum\limits_{k,p=0}^2 \hat{G}_{kp} dX_k dX_p \,,
\ee
where the following quantities have been defined  
\begin{equation} 
	\left( g_{ij} \right) =
	\left(\begin{matrix}
		G_{\Theta\Theta}& 0 \\
		0 & G_{yy}
	\end{matrix}\right)\,,\qquad i,j=1,2\,,
\end{equation}
\begin{equation}
	\left(\hat{G}_{kp}\right)=\left(\begin{matrix}
		G_{TT} & G_{T\Phi} & G_{T\Psi}\\
		G_{T\Phi}& G_{\Phi\Phi} & G_{\Phi\Psi}\\
		G_{T\Psi} & G_{\Phi\Psi} & G_{\Psi\Psi}
	\end{matrix}\right)\,, \qquad k,p=0,1,2\,,
\end{equation}
and the submatrix 
\begin{equation}
	\left( \hat{g}_{kp} \right)=\left(\begin{matrix}
		G_{\Phi\Phi} & G_{\Phi\Psi}\\
		G_{\Phi\Psi} & G_{\Psi\Psi}
	\end{matrix}\right)\,, \qquad k,p=1,2\,.
\end{equation}
We note, that for the inverse of $\left( \hat{g}_{kp} \right)$ matrix, we find
\begin{equation}
	\left( \hat{g}^{kp} \right)=\left(\begin{matrix}
		\frac{y^4\Xi^3+2a^2 M \cos^2\Theta}{y^2 \sin^2\Theta\,(y^4\Xi^3+2a^2 M )} & -\,\frac{2a^2 M}{y^2\,(y^4\Xi^3+2a^2 M )}\\
		-\,\frac{2a^2 M}{y^2\,(y^4\Xi^3+2a^2 M )}& \frac{y^4\Xi^3+2a^2 M \sin^2\Theta}{y^2 \cos^2\Theta\,(y^4\Xi^3+2a^2 M )} 
	\end{matrix}\right)\,, \qquad k,p=1,2\,.
\end{equation}

The components of the induced metric on the worldsheet: 
\begin{multline}\label{ws-metric}
	ds^2 _{ws}=  \left(\sum\limits_{i,j=1}^2 g_{ij}\,\dot{\xi}_i\dot{\xi}_j + \sum\limits_{k,p=0}^2\hat{G}_{kp}\,\dot{x}_k\dot{x}_p \right) d\tau^2 + \left(\sum\limits_{k,p=0}^2\hat{G}_{kp}\,m_k m_p\right)\, d\sigma^2 +\\
	+2\left(\sum\limits_{k,p=0}^2\hat{G}_{kp}\, m_k\dot{x}_p\right)\, d\tau d\sigma\,.
\end{multline}
The Nambu-Goto action \eqref{SNGg} with (\ref{indm-h}) becomes
%
%
%
\begin{multline} \label{NG-ws}
	S_{NG}= -\,\dfrac{1}{\alpha^\prime }\int\, d\tau\,\sqrt{ \left(\sum\limits_{k,p=1}^2\hat{G}_{kp}\, m_k\dot{x}_p  + \sum\limits_{k=1}^2 \hat{G}_{k0}\, m_k\, \kappa \right)^2+\,\,}\\ 
	\overline{\,\,-  \left\|\vec{m}\right\|^2\,\left(\sum\limits_{i,j=1}^2 g_{ij}\,\dot{\xi}_i\dot{\xi}_j + \sum\limits_{k,p=1}^2\hat{g}_{kp}\,\dot{x}_k\dot{x}_p + \hat{G}_{00}\,\kappa^2 +2\, \sum\limits_{k=1}^2\hat{G}_{0k}\,\kappa\dot{x}_k    \right)  \,\,}   \,,
\end{multline}
where $1/\alpha ^ \prime =\sqrt{\lambda}$ is the 't Hooft coupling constant and 
\begin{align}\label{A2}
	\left\|\vec{m}\right\|^2 =
	\sum\limits_{k,h=0} ^2 \hat{G}_{kh} m_k m_h =\sum\limits_{k,h=1} ^2 \hat{G}_{kh}(\xi_1,\xi_2) m_k m_h\equiv\sum\limits_{k,h=1} ^2 \hat{g}_{kh}(\xi_1,\xi_2) m_k m_h\,>\,0\,.
\end{align}
The problem reduces again to the dynamics of an effective point-particle with Lagrangian
\begin{multline} \label{NG-ws}
	L_{eff}= -\,\sqrt{\lambda}\,\sqrt{ \left(\sum\limits_{k,p=1}^2\hat{g}_{kp}\, m_k\dot{x}_p  + \sum\limits_{k=1}^2 \hat{G}_{k0}\, m_k\, \kappa \right)^2  -  \left\|\vec{m}\right\|^2\,\left(\sum\limits_{i,j=1}^2 g_{ij}\,\dot{\xi}_i\dot{\xi}_j + \,\,\right. }\\
	\overline{\left. \,\,+\sum\limits_{k,p=1}^2\hat{g}_{kp}\,\dot{x}_k\dot{x}_p + \hat{G}_{00}\,\kappa^2 +2\, \sum\limits_{k=1}^2\hat{G}_{0k}\,\kappa\dot{x}_k    \right)  \,\, }   \,.
\end{multline}
For convenience we will use the shorthand notation
\begin{multline} 
	\sqrt{ \left(\sum\limits_{k,p=1}^2\hat{g}_{kp}\, m_k\dot{x}_p  + \sum\limits_{k=1}^2 \hat{G}_{k0}\, m_k\, \kappa \right)^2  -  \left\|\vec{m}\right\|^2\,\left(\sum\limits_{i,j=1}^2 g_{ij}\,\dot{\xi}_i\dot{\xi}_j + \,\,\right. }\\
	\overline{\left. \,\,+\sum\limits_{k,p=1}^2\hat{g}_{kp}\,\dot{x}_k\dot{x}_p + \hat{G}_{00}\,\kappa^2 +2\, \sum\limits_{k=1}^2\hat{G}_{0k}\,\kappa\dot{x}_k    \right)  \,\, }\,\equiv\, \sqrt{......\,\,}   \,.
\end{multline}
As before, it is useful to consider the  Hamiltonian formulation of the problem. To this end, we have to calculate the canonical momenta
\begin{align}\label{momenta}
	\Pi_{\xi_i}&=\frac{\partial L_{eff}}{\partial \dot{\xi}_i} =\sqrt{\lambda} \,\dfrac{ \left\|\vec{m}\right\|^2 g_{ij}\,\dot{\xi}_j } {\sqrt{......\,\,}} \,, \\
	\Pi_{x_p}&=\frac{\partial L_{eff}}{\partial \dot{x}_p} =\sqrt{\lambda}\, \dfrac{ \,\left\|\vec{m}\right\|^2 \hat{g}_{pq}\, \dot{x}_q - (\hat{g}_{kq}\, m_k\dot{x}_q)\, \hat{g}_{pq}\, m_q +\left\|\vec{m}\right\|^2 \hat{G}_{p0}\,\kappa -(\hat{G}_{k0}\, m_k\kappa)\, \hat{g}_{pq}\, m_q\,}
	{\sqrt{......\,\,}}\,,
\end{align}
which also implies the constraint 
\begin{equation}\label{constraintPi}
	\sum\limits_{p=1}^2m_p\, \Pi_{x_p}=0.
\end{equation}
 Applying a Legendre transformation, it is straightforward to find the (square of) the pulsating string Hamiltonian
\begin{multline}\label{Ham^2}
	H^2= \kappa^2 \, \left(\sum\limits_{l,s=1}^2 \hat{G}_{0l}\,\hat{g}^{ls}\, \Pi_{x_s}\right)^2 +\\
	- \kappa^2 \,\left( \hat{G}_{00} +  \sum\limits_{l,s=1}^2 \hat{G}_{0l}\,\hat{g}^{ls}\, \hat{G}_{s0} \right) \,\left\lbrace          \sum\limits_{i,j=1}^2   \Pi_{\xi_i}\,g^{ij} \,\Pi_{\xi_j}   +   \sum\limits_{i,j=1}^2   \Pi_{x_i}\,\hat{g}^{ij} \,\Pi_{x_j} +\lambda\,\left\|\vec{m}\right\|^2 \right\rbrace \,.
\end{multline}
The explicit expressions for the terms in the brackets are
\begin{equation}
	\sum\limits_{l,s=1}^2 \hat{G}_{0l}\,\hat{g}^{ls}\, \Pi_{x_s}\,=\,-\, h_2(y)\,\left(  \Pi_{x_1}+\Pi_{x_2} \right) 
\end{equation}
and
\begin{equation}
	\hat{G}_{00} +  \sum\limits_{l,s=1}^2 \hat{G}_{0l}\,\hat{g}^{ls}\, \hat{G}_{s0} \,=\,-\,\frac{y^2 \Xi^3 -2aM\,h_1(y)\,h_2(y) }{h_1(y)\,y^2 \Xi^3}\,\equiv\,-\,K^2(y)\,,
\end{equation}\,
where
\begin{equation}\label{funcsh1h2}
	h_1(y)\,=\,\frac{y^{2}\Xi^{3}}{-2M + y^{2}(1+\ell^{2}y^{2})\Xi^{3}}\,, \qquad h_2(y)\,=\,\frac{2aM}{2a^{2}M + y^{4}\Xi^3}\,.
\end{equation}
The expression of $H^2$ can be conveniently written 
%
%
%
\begin{equation}\label{Ham^2 2}
	\frac{H^2}{\kappa^2}\,=\,K^2(y)\left\lbrace   \sum\limits_{i,j=1}^2   \Pi_{\xi_i}\,g^{ij} \,\Pi_{\xi_j} + \sum\limits_{i,j=1}^2   \Pi_{x_i}\,\left[  \frac{h_2 ^2 (y)}{K^2(y)}\,\hat{\delta}^{ij} +\hat{g}^{ij} \right] \,\Pi_{x_j} +\lambda\,\left\|\vec{m}\right\|^2     \right\rbrace \,,
\end{equation}
where
\begin{equation}
\left(\hat{\delta}^{ij} \right)\,=\,\left(\begin{matrix}
	1& 1 \\
	1& 1
\end{matrix}\right)\,,\,\, i,j=1,2 	\,.
\end{equation}
It should be noted here that the above Hamiltonian \eqref{Ham^2} can be considered as the effective Hamiltonian of an effective point-particle on the Kerr-AdS background. The last term 
\begin{equation}
	-\lambda\,\kappa^2\,\left\|\vec{m}\right\|^2 \,\left( \hat{G}_{00} +  \sum\limits_{l,s=1}^2 \hat{G}_{0l}\,\hat{g}^{ls}\, \hat{G}_{s0} \right)\,\equiv\, \lambda\,\kappa^2\,\left\|\vec{m}\right\|^2 \,K^2(y)\,\equiv\,\lambda\,U(\Theta,\,y)\,,
\end{equation}
serves as an effective potential, which encodes the relevant dynamics of the strings.\\
In the context of the holographic correspondence, the potential is very small comparing to the kinetic part. Therefore, we can calculate quantum corrections to the energy by making use of the perturbation theory. 

\subsubsection{Laplace-Beltrami operator and wave function}
The standard Laplace-Beltrami operator in the AdS coordinates for the 5d Kerr-AdS \eqref{GLKA} is
\begin{multline}
	\bigtriangleup^{(5)}_{Kerr-AdS}\,=\,G^{TT}\frac{\partial^2}{\partial T^2}+2\,G^{T\Phi(\Psi)}\frac{\partial}{\partial T}\left(\frac{\partial}{\partial \Phi}+\frac{\partial}{\partial \Psi} \right) +G^{\Phi\Phi}\frac{\partial^2}{\partial \Phi^2}+2\,G^{\Phi\Psi}\frac{\partial^2}{\partial \Phi\partial\Psi}+G^{\Psi\Psi}\frac{\partial^2}{\partial \Psi^2}+\\
	+\frac{1}{\sqrt{-\det G}}\,\frac{\partial}{\partial \Theta}\left(\sqrt{-\det G\,}\,G^{\Theta\Theta}\,\frac{\partial}{\partial \Theta} \right)+ +\frac{1}{\sqrt{-\det G}}\,\frac{\partial}{\partial y}\left(\sqrt{-\det G\,}\,G^{yy}\,\frac{\partial}{\partial y} \right)\,.
\end{multline}
In the notations of Appendix A (see \eqref{inverseG} and \eqref{detG}), we have
\begin{multline}\label{Laplace}
	\bigtriangleup^{(5)}_{Kerr-AdS}\,=\,-h_1(y)\,\frac{\partial^2}{\partial T^2}-2\,h_1(y)\,h_2(y)\,\frac{\partial}{\partial T}\left(\frac{\partial}{\partial \Phi}+\frac{\partial}{\partial \Psi} \right) +\\
	-\,\frac{1}{y^2}a\,h_2(y)\,\left(\frac{\partial}{\partial \Phi}+\frac{\partial}{\partial \Psi} \right)^2 +  \frac{1}{y^2\sin^2\Theta} \,\frac{\partial^2}{\partial \Phi^2}+\frac{1}{y^2\cos^2\Theta} \,\frac{\partial^2}{\partial \Psi^2}+\\
	+\frac{1}{y^2}\,\frac{1}{\sin(2\Theta)}\frac{\partial}{\partial \Theta}\left(\sin(2\Theta)\,\frac{\partial}{\partial \Theta} \right)+ \frac{1}{\sqrt{h(y)\,}}\,\frac{\partial}{\partial y}\left(\sqrt{h(y)\,}\,G^{yy}\,\frac{\partial}{\partial y} \right)\,,
\end{multline}
where $h_1(y)$ and $h_2(y)$ are given by (\ref{funcsh1h2}).

To obtain the wave function, we have to solve the equation
\begin{equation}\label{kernel}
	\bigtriangleup^{(5)}_{Kerr-AdS}\,F(T,\,\Theta,\,y,\,\Phi,\,\Psi)\,=\,0\,,
\end{equation}
using separation of variables
\begin{equation}\label{separate}
	F(T,\,\Theta,\,y,\,\Phi,\,\Psi)\,=\,e^{-i E\, T}\, e^{i n_1\Phi}\, e^{i n_2\Psi}\, f(\Theta,\,y)\,,\qquad n_1,\,n_2\, \in\mathbb{Z}\,.
\end{equation}
Substituting \eqref{separate} in to the equation \eqref{kernel}, we find the following partial differential equation for $\, f(\Theta,\,y)\,$
\begin{multline}\label{laplaceEquation}
	E^2\,h_1(y)\,f-2h_1(y)h_2(y)\,E\,(n_1+n_2)\,f+\frac{1}{y^2}\,ah_2(y)\,(n_1+n_2)^2\,f+\\
	-\frac{1}{y^2}\,\left(  \frac{n_1^2}{\sin^2\Theta} +\frac{n_2^2}{\cos^2\Theta}\right) \,f        +\frac{1}{y^2}\,\frac{1}{\sin(2\Theta)}\frac{\partial}{\partial \Theta}\left(\sin(2\Theta)\,\frac{\partial f}{\partial \Theta} \right)+\\
	+ \frac{1}{\sqrt{h(y)\,}}\,\frac{\partial}{\partial y}\left(\sqrt{h(y)\,}\,G^{yy}\,\frac{\partial f}{\partial y} \right) \,=\,0\,.
\end{multline}
We can further separate the variables defining $\, f(\Theta,\,y)\,=\,f(\Theta)\,Y(y)\,$.
\subsubsection{Solving the Schr\"odinger equation on the reduced subspace $y=const$}

In the subsection (3.2), we obtain an explicit pulsating solution \eqref{puls solution} in the case for the string winding numbers satisfy $\,m_1^2=m_2^2\equiv\,m^2\,$ and  $\,y=const\neq\,0\,$.\\
In this case, taking into account \eqref{constraintPi}, for the first term of the Hamiltonian \eqref{Ham^2}, we have
\begin{equation}
	\left( \sum\limits_{l,s=1}^2 \hat{G}_{0l}\,\hat{g}^{ls}\, \Pi_{x_s}\right) ^2\,=\,h^2_2(y)\,\left(  \Pi_{x_1}+\Pi_{x_2} \right)^2\,=\,0\,. 
\end{equation}
Moreover, all functions of $y$ are constants and 
\begin{equation}
	\hat{G}_{00} +  \sum\limits_{l,s=1}^2 \hat{G}_{0l}\,\hat{g}^{ls}\, \hat{G}_{s0} \,=\,-\frac{y^2 \Xi^3 -2aM\,h_1(y)\,h_2(y) }{h_1(y)\,y^2 \Xi^3}\,\equiv\,-K^2 =const\,,
\end{equation}
where $h_{1}$ and $h_{2}$ \eqref{funcsh1h2} are taken to be constant.
%
%
Therefore, the square of the Hamiltonian \eqref{Ham^2}  has the form 
\begin{equation}\label{Ham3}
	H^2=\,\kappa^2\,K^2\,\left\lbrace   \Pi_{\Theta}\,g^{\Theta\Theta} \,\Pi_{\Theta}   +   \sum\limits_{i,j=1}^2   \Pi_{x_i}\,\hat{g}^{ij} \,\Pi_{x_j} +\lambda\,\left\|\vec{m}\right\|^2 \right\rbrace \,.
\end{equation}
We observe that $H^2$ looks like a point-particle Hamiltonian, which seems to be characteristic feature for pulsating strings in holography. The last term, 
\begin{equation}\label{potentialTheta}
	\lambda\,\kappa^2\,K^2\left\|\vec{m}\right\|^2 \,\equiv\,\lambda\,U(\Theta)
\end{equation}
serves as an effective potential, which encodes the relevant dynamics of the strings.\\
In the context of holographic correspondence, the potential is very small compared to the kinetic part. Therefore, we can calculate quantum corrections to the energy using perturbation theory.
%
\paragraph{Wave functions}
%
The kinetic term of the Hamiltonian \eqref{Ham3} can be considered as a three dimensional Laplace-Beltrami operator of the Kerr-AdS subspace with $y=const$
\begin{equation}
	\vec{P}^2\,=\,\left\lbrace   \Pi_{\Theta}\,g^{\Theta\Theta} \,\Pi_{\Theta}   +   \sum\limits_{i,j=1}^2   \Pi_{x_i}\,\hat{g}^{ij} \,\Pi_{x_j} \right\rbrace \,\,\longrightarrow \,\, \bigtriangleup^{(3)}_{Kerr-AdS}\,,
\end{equation}
which defines the eigen-functions of the Hamiltonian, satisfying the following Schr\"odinger equation
\begin{equation}\label{Schro}
	\bigtriangleup^{(3)}_{Kerr-AdS}\,F\,=\,-\,\frac{E^2}{\kappa^2\,K^2}\,F\,.
\end{equation}
Comparing it with the equation \eqref{laplaceEquation}, we notice that it corresponds to taking $n_1=-n_2\equiv\,n\,,\,\, n\in\mathbb{Z}\,$. Thus, we can write the equation \eqref{Schro} in the form
	\footnote{In this case, the solution of the below equation can be written directly in terms of Shifted Legendre polynomials, but below we will follow the more general procedure.}
\begin{equation}\label{equation}
	\left\lbrace \frac{1}{y^2}\,\frac{1}{\sin\Theta\,\cos\Theta}\,\frac{\partial}{\partial\Theta}\left( \sin\Theta\,\cos\Theta\,\frac{\partial}{\partial\Theta}\right) -\frac{n^2}{y^2}\,\frac{1}{\sin^2\Theta\,\cos^2\Theta}+\frac{E^2}{\kappa^2\,K^2}\right\rbrace \,F(\Theta)\,=\,0\,.
\end{equation}
 Since in this case the potential $\,\lambda\,U$ is a constant, the corresponding Schr\"odinger equation can be easily solved. The eigenvalue problem for the Hamilton (square of) operator is exactly solvable.\\
The last term in the equation \eqref{equation} is shifted with a constant and can be written as
\begin{equation}\label{Legendre}
	\left\lbrace \frac{1}{\sin\theta}\,\frac{\partial}{\partial\theta}\left( \sin\theta\,\frac{\partial}{\partial\theta}\right) -\frac{n^2}{\sin^2\theta}+\frac{y^2\,(E_{full}^2 +\lambda\,U)}{4\,\kappa^2\,K^2}\right\rbrace \,F(\theta)\,=\,0\,,\qquad \theta\equiv\,2\Theta\,.
\end{equation}
As far as, we want square integrable eigenfunctions with a discrete spectrum, the solution of the above equation can be directly written in terms of Shifted Legendre polynomials
\begin{equation}
	F(\theta)\,\sim\, P_n ^k (\cos\theta)\,, \,\, -k\,\leq\,n\,\leq\,k\,,\,\,k\in\mathbb{N}\,.
\end{equation}
The discrete spectrum is determined by
\begin{equation}
	\frac{y^2\,(E_{full}^2 +\lambda\,U)}{4\,\kappa^2\,K^2}\,=\,k(k+1)\,.
\end{equation}
Below we will follow slightly more general procedure, which is also valid for non-constant potentials.\\
To this end it is convenient to define a new variable $z=\sin^2\Theta\,, \,\, 0\leq\,z\,\leq\,1\,$. Then the equation \eqref{equation} can be written as
\begin{equation}\label{eq_z}
	\left\lbrace  \frac{d^2}{dz^2} +\frac{(1-2z)}{z(1-z)} \,\frac{d}{dz} - \frac{N^2}{z^2(1-z)^2} + \frac{\hat{E}^2}{z(1-z)}  \right\rbrace \,F(z)\,=\,0\,,
\end{equation}  
where $\,N=\frac{n^2}{4}\,$ and $\,\hat{E}^2\,=\,\frac{y^2\,E^2}{\kappa^2\,K^2}\,$.\\
The general solution is the following linear combination
\begin{multline}
	F(z)=C_1\,z^N\,(1-z)^N\,{_2F_1}\left[   2N +\frac{1}{2}+\frac{1}{2}\,\sqrt{1+4\hat{E}^2}\,,\,   2N +\frac{1}{2}-\frac{1}{2}\,\sqrt{1+4\hat{E}^2}\,;\, 1+2N\,;\,z \right] + \\
	+C_2\,z^{-N}\,(1-z)^N\,{_2F_1}\left[  \frac{1}{2}+\frac{1}{2}\,\sqrt{1+4\hat{E}^2}\,,\,   \frac{1}{2}-\frac{1}{2}\,\sqrt{1+4\hat{E}^2}\,;\, 1-2N\,;\,z \right] \,.
\end{multline}
Since the second term is singular at zero, we set $\,C_2 =0\,$, and the solution satisfying the boundary conditions is 
\begin{equation}\label{solution}
	F(z)=C\,z^N\,(1-z)^N\,{_2F_1}\left[   2N +\frac{1}{2}+\frac{1}{2}\,\sqrt{1+4\hat{E}^2}\,,\,   2N +\frac{1}{2}-\frac{1}{2}\,\sqrt{1+4\hat{E}^2}\,;\, 1+2N\,;\,z \right]\,.
\end{equation}
In addition, we have to ensure that the solutions $F(\Theta)$ are square integrable with respect to the measure $\Theta$ (respectively $z$). The integrability condition leads to the following restriction on the parameters
\begin{equation}\label{quantE}
	2N +\frac{1}{2}-\frac{1}{2}\,\sqrt{1+4\hat{E}^2}\,=\,-k\,,\qquad k\in\mathbb{N}\,.
\end{equation}
This requirement imposes energy quantization:
\begin{equation}\label{E^2}
	E^2\,=\,\kappa^2\,\frac{K^2}{4y^2}\,\left[ (4N+1+k)^2 -1 \right].
\end{equation}
The condition \eqref{quantE} converts the solution \eqref{solution} in terms of Jacobi ortogonal polynomials (in this case the solution of the equation can also be written directly in terms of Shifted Legendre polynomials)
\begin{equation}\label{Jacobi z}
	F(z)=\,C\,z^{\alpha/2}\,(1-z)^{\beta/2}\,\dfrac{k!\,\Gamma(\alpha+1)}{\Gamma(\alpha+1+k)}P^{(\alpha,\beta)}_{k}(1-2z)\,,\qquad k\in\mathbb{N}\,,
\end{equation}
where $ \,\alpha=\beta\,=\,2N\equiv\,\frac{n^2}{2} \,,\,\, n\in\mathbb{Z}\,$.
It is more convenient to work in terms of  $\,u \equiv\,1-2z\,,\,\,-1\leq\,u\,\leq\,1\,$
\begin{equation}\label{Jacobi u}
	F_{k,n}(u)=\,C\,\left(  \frac{1-u}{2} \right) ^{\alpha/2}\,\left(  \frac{1-u}{2} \right) ^{\alpha/2}\,\dfrac{k!\,\Gamma(\alpha+1)}{\Gamma(\alpha+1+k)}P^{(\alpha,\beta)}_{k}(u)\,,\,\,-1\leq\,u\,\leq\,1\,\,.
\end{equation}
 Then with respect to the measure
\begin{equation}\label{measure}
	d\Omega =\,\sqrt{-\det G^{(4)}}\,d\Theta\,d\Phi\,d\Psi\,=\,-\,\sqrt{\frac{h_1\,h_2}{2 a M \Xi^3 \,y^2}}\,\,\frac{du}{4}\,d\Phi\,d\Psi,
\end{equation}
we find that the normalized  wave function is
\begin{equation}\label{wave f}
	f_{k,n}(u)=\,\sqrt{\frac{(2\alpha+1+2k)\,k!\,\Gamma(2\alpha+1+k)}{2^{\alpha-1}\,
			\Gamma(\alpha+1+k)\,\Gamma(\alpha+1+k)}} \,(1-u)^{\alpha/2}\,(1+u)^{\alpha/2}P^{(\alpha,\alpha)}_{k}(u)\,.
\end{equation}
Finally, the total free wave functions have the form
\begin{align}\label{tot wave f}
	f^{tot}_{k,n}(u,\,\Phi,\,\Psi)=\,&\sqrt{\frac{(2\alpha+1+2k)\,k!\,\Gamma(2\alpha+1+k)}{\,\omega(y)\,2^{\alpha-1}\,
			\Gamma(\alpha+1+k)\,\Gamma(\alpha+1+k)}}\,\times\nonumber\\
	&\times\,(1-u)^{\alpha/2}\,(1+u)^{\alpha/2}P^{(\alpha,\alpha)}_{k}(u)\,e^{in\Phi}\,e^{-in\Psi}\,.
\end{align}
The next step is to calculate perturbatively the corrections to the energy of the free ground states.
	\paragraph{Leading correction to the energy}
	Perurbatively, the first correction to the energy reads
	\begin{equation}\label{delta-E}
		\delta E^2=\,\lambda\,\langle\,f^{tot}_{k,n}\,\vert\,U\,\vert\,f^{tot}_{k,n}\,\rangle
		=\lambda \int\limits_{\!\!\!\!\!-1}^1\!  \int\limits_0^{2\pi}\! \int\limits_0^{2\pi}\, \left| f^{tot}_{k,n}(u,\,\Phi,\,\Psi) \right|^2\,\, U(u,\,\Phi,\,\Psi)\,\, d\Omega(u,\, \Phi,\,\Psi) \,.
	\end{equation}
	Let us remind the form of the potential \eqref{potentialTheta}
	\begin{equation}\label{potential}
		U(\Theta)\,=\,\kappa^2\,K^2\left\|\vec{m}\right\|^2 \,=\,\kappa^2\,K^2\,\sum\limits_{k,h=1} ^2 \hat{g}_{kh}(y,\,\Theta) m_k m_h\,.
	\end{equation}
	Since for $\,y=const\,$, the value of $\,\left\|\vec{m}\right\|^2\,$ is also a constant, the potential is
	\begin{equation}\label{potential costant}
		U\,=\,\kappa^2\,K^2 \,m^2\, \left( y^2 + \frac{2a^2 M}{y^2\Xi^3}\right)=const\,.
	\end{equation}
	Using the scalar product \eqref{delta-E}, one can easily compute the first correction to the energy
	\begin{equation}\label{delta-E value}
		\delta E^2=\,\lambda\,\langle\,f^{tot}_{k,n}\,\vert\,U\,\vert\,f^{tot}_{k,n}\,\rangle
		=\,\lambda\,U\, \langle\,f^{tot}_{k,n}\,\vert\,f^{tot}_{k,n}\,\rangle\,=\,\lambda\,U\,=\, \kappa^2\,K^2 \,m^2\, \left( y^2 + \frac{2a^2 M}{y^2\Xi^3}\right)   \,.
	\end{equation}
	%
	According to the standard holographic dictionary, the anomalous dimension of the corresponding dual operators are directly related to the corrections of the string energy.  The interpretation of results from the holographic point of view is not straightforward since the dual theory is at finite temperature. Nevertheless, near or at conformal point the expressions can be thought of as the dispersion relations of stationary states.\\
	%

	\newpage
	\section{Conclusions} \label{Sect5}
	
In this paper, we have explored pulsating strings in the 5d Kerr-AdS black hole. 
	The holographic dual for the  Kerr-$AdS_{5}$ black hole is  $\mathcal{N}=4$ SYM on $\mathbb{S}^{1}\times \mathbb{S}^{3}$ at finite temperature (a thermal ensemble).
	 For simplicity we have focused on  the Kerr-$AdS_{5}$ background 	with equal rotating parameters in the static-at-infinity frame.
	 We have found exact solutions describing pulsating strings in the Kerr-$AdS_{5}$ background. 
	 To construct this we have reduced the string Lagrangian to an effective model of a mechanical particle with a potential. 
	 We have shown that the corresponding equations of motions can be solved in terms of quadratures. 
	 The periodicity of the string solutions are guaranteed  by certain constraints on the parameters of the original model.
	 Thus the pulsating string solutions oscillate between two turning points, which are above the horizon.
	It worth to be noted that, these solution  are the first pulsating string  solutions in the Kerr-$AdS_{5}$ black holes.
	 
	 Moreover, we have computed the energy of the pulsating string the Kerr-$AdS_{5}$ background  following the Bohr-Sommerfeld analysis. 
	 We have found  corrections related to the rotation and temperature of the black hole. At zero value of the rotational parameter the relation for the energy tends to be known from \cite{Alishahiha:2002fi}.

	
	In this paper we picked the asymptotically AdS black hole in five dimensions, Kerr-AdS, which possesses $SO(4)$ symmetry. If the space-time is only asymptotically anti-de Sitter it corresponds to UV conformal fixed points in the boundary theory.  Compared to AdS case  the IR behavior is expected to be quite different, with the horizon now playing the role of a thermal background.
	The holographic interpretation of the space-time conserved quantities however is not unique. It essentially depends on the choice of the conformal structure of the asymptotic metric at the  boundary. As we already mentioned in the bulk text, the notion of anomalous dimensions is well defined only in the vicinity of the conformal point.  This makes the interpretation from holographic point of view somewhat complicated and subtle. One way to make sense of the expressions is to consider two-point function, or scalar bulk-to-bulk Green’s functions and see its behavior when one of the insertions is approaching the boundary. Indeed, one can see the considerations near conformal point are quite reasonable, thus the expressions can be interpreted as dispersion relations of stationary states in the gauge theory side.

	 
	 Thinking about more general picture of the Kerr-AdS/CFT correspondence, many issues remain to be investigated and understood. In this paper we made some approximations to find the corrections to the string energy. It will be also interesting to consider other holographic observables, i.e. thermal $n$-point correlation functions, in the Kerr-$AdS_{5}$ background as it was done for non-rotating AdS black holes \cite{Kraus:2017kyl}-\cite{Karlsson:2021duj}. Particularly, to probe holographic four-point functions one needs to study a motion of a highly energetic particle in the 5d Kerr-AdS black hole \cite{Kulaxizi:2018dxo}-\cite{Kulaxizi:2019tkd}. 
	 
	 Kerr-AdS black holes are remarkable with that its wave equation allows separation of variables. The radial and angular equations are of Fuchsian type and analyzing structure of singularities can be reduced to Heun type. Thus, it is literally calling to apply method from integrable systems and Riemann-Hilbert problem. Indeed,  the authors of \cite{Litvinov:2013sxa} have shown that the monodromy problem of the Heun equation is related to the connection problem for the Painleve VI and conformal blocks in 2d CFT, in particular Liouville field theory. Applying such an approach to scattering off black holes the authors of \cite{Novaes:2014lha}  have found that for generic charges the problem can be reduced to the Painleve VI transcendent. Moreover, the accessory parameters are expressed through the charges. 
	For the 5d case of the Kerr-AdS black holes scalar perturbations  it was shown that corrections to the extremal limit can be encoded in the monodromy parameters of the Painlevé V transcendent \cite{Amado:2017kao}-\cite{BarraganAmado:2021uyw}. 
	 Pulsating strings approach also relates dispersion relations to conserved charges, as well as field theory  characteristics.  It would be very interesting to investigate how and why all these quatities are related. Is there any relation to Alday-Gaiotto-Tachikawa conjecture? 
	 	
	 	We hope to return to these issues in the near future.	 
		 	
	
	\paragraph{Acknowledgements} \
	\\
	A.G., R.R. and H.D. would like to thank Alexey Isaev and Sergey Krivonos for insightful discussions. A.G. is also grateful to I.Ya. Aref'eva for useful discussions. H. D. would like also to thank T. Vetsov for discussions on various issues of holography. The work of R.R. and H.D. is partially supported by the Program “JINR– Bulgaria” at Bulgarian Nuclear
	Regulatory Agency. The work of R.R. and H.D. was also supported in part by BNSF H-28/5. The work of A.G. is supported by Russian Science Foundation grant 20-12-00200.
	
\newpage
\appendix
\section{The geometric characteristics of the metric}
\subsection{Non-zero mertic components of Kerr-$AdS_{5}$}
The metric components of the Kerr-AdS black hole metric  \eqref{GLKA}  are given by
	\bea\label{App}
	G_{TT}&=& - (1 + y^{2}l^{2}-\frac{2M}{y^{2}\Xi^{3}}),\quad G_{yy}=\frac{y^{4}}{y^{4}(1+y^{2}\ell^{2})-\frac{2M}{\Xi^{2}}y^{2} + \frac{2Ma^{2}}{\Xi^{3}}},\quad G_{\Theta\Theta}=y^{2},\nonumber\\
	G_{\Phi\Phi}&=&\sin^{2}\Theta\left(y^{2}+\frac{2a^{2}M}{y^{2}\Xi^{3}}\sin^{2}\Theta\right),\quad
	G_{\Psi\Psi}=\cos^{2}\Theta\left(y^{2} +\frac{2a^{2}M\cos^{2}\Theta}{y^{2}\Xi^{3}}\right),\\
	G_{T\Phi}&=&G_{\Phi T}=-\frac{2aM\sin^{2}\Theta}{y^{2}\Xi^{3}},\quad G_{T\Psi}=G_{\Psi T}=-\frac{2aM \cos^{2}\Theta}{y^{2}\Xi^{3}},\nonumber\\ \label{App2}
	G_{\Phi\Psi}&=&G_{\Psi\Phi}=\frac{2Ma^{2}\sin^{2}\Theta \cos^{2}\Theta}{\Xi^{3}y^{2}}.\nonumber
	\eea

The inverse non-zero components of the 5d Kerr-AdS metric \eqref{GLKA} 
\bea
G^{TT}& =& - \frac{y^{2}\Xi^{3}}{-2M + y^{2}(1+\ell^{2}y^{2})\Xi^{3}}, \quad G^{\Theta\Theta} = \frac{1}{y^{2}},\nonumber \\
G^{T\Phi}& =&\frac{2aMy^{2}\Xi^{3}}{(2a^{2}M + y^{4}\Xi^3)(2M -y^{2}(1+\ell^2 y^2)\Xi^{3})}\,=\,G^{T\Psi},\nonumber \\
G^{\Phi\Phi}& =&\frac{1}{y^{2}}\left(-\frac{2a^{2}M}{2a^2 M + y^4 \Xi^3} + \frac{1}{\sin^{2}\Theta}\right),\, G^{\Psi\Psi}= \frac{1}{y^{2}}\left(-\frac{2a^{2}M}{2a^2 M + y^4 \Xi^3} + \frac{1}{\cos^{2}\Theta}\right),\nonumber\\
G^{\Phi\Psi} &=& -\frac{2a^{2}M}{2a^{2}My^{2} + y^{6}\Xi^{3}},\quad G^{yy} = 1 + \ell^2 y^2 + \frac{2M(a^{2} - y^{2}\Xi)}{y^{4}\Xi^{3}}\,.
\eea
The determinant of the metric reads
\bea
\det G =  \frac{y^{4}(2a^{2}M + y^{4}\Xi^{3})(2M - y^{2}(1+ \ell^2y^{2})\Xi^{3})\sin^{2}(2\Theta)}{4\Xi^{3}(2a^{2}M - 2My^{2}\Xi + y^{4}(1+ \ell^{2}y^{2})\Xi^3)}\,.
\eea
In terms  $\,\,h_1(y)\,=\,\frac{y^{2}\Xi^{3}}{-2M + y^{2}(1+\ell^{2}y^{2})\Xi^{3}}\,, \,\,h_2(y)\,=\,\frac{2aM}{2a^{2}M + y^{4}\Xi^3}\,\,$ we can rewrite the inverse non-zero components of the 5d Kerr-AdS metric 
\bea \label{inverseG}
G^{TT}& =& - h_1(y)\,, \quad G^{\Theta\Theta} = \frac{1}{y^{2}} \,,\nonumber\\
G^{T\Phi}& =&-h_1(y)\,h_2(y)=G^{T\Psi}\,, \\
G^{\Phi\Phi}& =& \frac{1}{y^{2}}\left(-a\,h_2(y) + \frac{1}{\sin^{2}\Theta}\right), \quad G^{\Psi\Psi}= \frac{1}{y^{2}}\left(-a\,h_2(y) + \frac{1}{\cos^{2}\Theta}\right)\,\nonumber\\
G^{\Phi\Psi} &=& -\frac{1}{y^2}\,a\,h_2(y)\,,\quad G^{yy} = 1 + \ell^2 y^2 + \frac{2M(a^{2} - y^{2}\Xi)}{y^{4}\Xi^{3}}\,\nonumber
\eea
and the determinant of the metric reads
\begin{equation}\label{detG}
	\det G=\,-\,\frac{aM\,y^6}{2h_1(y)\,h_2(y)\,\left[  2a^{2}M - 2My^{2}\Xi + y^{4}(1+ \ell^{2}y^{2})\Xi^3  \right] }\,\sin^2 (2\Theta)\,
	\equiv\,-\,h(y)\,\sin^2 (2\Theta)\,.
\end{equation}

\subsection{Roots of the blackening function}

The horizon for the Kerr-$AdS_{5}$ black hole is defined as the greatest root to the equation
\be
1 + y^{2}\ell^{2} - \frac{2M}{\Xi^{2}y^2} + \frac{2Ma^{2}}{\Xi^{3}y^4} =0.
\ee
There are 6 roots for it, namely, 
{\footnotesize
\bea\label{rootbf1}
&&y_{+} = \sqrt{\frac{-\Xi \sqrt[3]{3 \sqrt{3}\ell^2 \sqrt{ M  \mathbf{q}}- \mathbf{p}}+(3 \sqrt{3}\ell^2 \sqrt{M  \mathbf{q}}- \mathbf{p})^{2/3}+6 \ell^2 M +\Xi ^2}{3\ell^2 \Xi \sqrt[3]{3 \sqrt{3}\ell^2 \sqrt{ M \mathbf{q}}- \mathbf{p} }}}\\
&&y_{2} = -\sqrt{\frac{-\Xi \sqrt[3]{3 \sqrt{3} \ell^2\sqrt{ M \mathbf{q}}- \mathbf{p}}+ (3 \sqrt{3} \ell^2 \sqrt{ M  \mathbf{q}}- \mathbf{p})^{2/3}+6 \ell^2 M +\Xi^2}{3\ell^2 \Xi \sqrt[3]{3 \sqrt{3} \ell^2  \sqrt{ M \mathbf{q}}- \mathbf{p}}}},\\ \label{rootbf3}
&&y_{1}^{*} = -\sqrt{\frac{-2 \Xi  \sqrt[3]{3 \sqrt{3} \ell^2\sqrt{M \mathbf{q}}- \mathbf{p}}-(1+i \sqrt{3})(3 \sqrt{3}\ell^2 \sqrt{ M \mathbf{q}}- \mathbf{p})^{2/3}+6 i (\sqrt{3}+i) \ell^2 M +i (\sqrt{3}+i) \Xi ^2}{6\ell^2 \Xi \sqrt[3]{3 \sqrt{3} \ell^2\sqrt{ M  \mathbf{q}}- \mathbf{p}}}},\nonumber\\
\,\\
&& y_{2}^{*} = \sqrt{\frac{-2 \Xi \sqrt[3]{3 \sqrt{3} \ell^2\sqrt{M \mathbf{q}}- \mathbf{p}}-(1+i \sqrt{3}) (3 \sqrt{3} \ell^2\sqrt{ M \mathbf{q}}-\mathbf{p})^{2/3}+6 i(\sqrt{3}+i) \ell^2 M+i (\sqrt{3}+i) \Xi ^2}{6\ell^2 \Xi \sqrt[3]{3 \sqrt{3} \ell^2 \sqrt{ M\mathbf{q}}- \mathbf{p}}}},\nonumber\\
&&\, \\
&&y_{3}^{*}= -\sqrt{\frac{i (\sqrt[3]{3 \sqrt{3}\ell^2 \sqrt{M  \mathbf{q}}- \mathbf{p}}+\Xi) ((\sqrt{3}+i) \sqrt[3]{3 \sqrt{3} \ell^2  \sqrt{M \mathbf{q}}-\mathbf{p}}-(\sqrt{3}-i) \Xi)-6(1+ i \sqrt{3}) \ell^2 M }{6\ell^2 \Xi \sqrt[3]{3 \sqrt{3}\ell^2  \sqrt{ M \mathbf{q}}- \mathbf{p}}}}, \nonumber\\
\,\\ \label{rootbf6}
&& y_{4}^{*} = \sqrt{\frac{i (\sqrt[3]{3 \sqrt{3}\ell^2 \sqrt{ M \mathbf{q}}-\mathbf{p}}+\Xi ) ((\sqrt{3}+i) \sqrt[3]{3 \sqrt{3}\ell^2 \sqrt{ M  \mathbf{q}}- \mathbf{p}}-(\sqrt{3}-i) \Xi)-6(1+ i \sqrt{3}) \ell^2 M }{6\ell^2 \Xi \sqrt[3]{3 \sqrt{3}\ell^2\sqrt{ M \mathbf{q}}- \mathbf{p}}}},
\eea}
where we define
\be
\mathbf{p} =27 a^2 \ell^4 M+9 \ell^2 M \Xi +\Xi ^3, \quad  \mathbf{q}=27 a^4 \ell^4 M+2 a^2 (9 \ell^2 M \Xi +\Xi ^3)-M (8 \ell^2 M+\Xi ^2)
\ee


\begin{thebibliography}{99}
	 	
	 	\bibitem{Witten}
		E. Witten, “Anti-de Sitter space and holography,” {\it Adv. Theor. Math. Phys.} 2 (1998) 253–291, [\href{http://arxiv.org/abs/hep-th/9802150}{{\tt arXiv:hep-th/9802150}}].
		\bibitem{Witten2}
		E. Witten, “Anti-de Sitter space, thermal phase transition, and confinement in gauge theories,” {\it Adv. Theor. Math. Phys.} 2 (1998) 505–532,  [\href{http://arxiv.org/abs/hep-th/9803131}{{\tt arXiv:hep-th/9803131}}].

		\bibitem{HHT}
		S.W.Hawking, C.J.Hunter and M.Taylor-Robinson,
		Rotation and the AdS/CFT correspondence,
		{\it Phys.Rev.} {\bf D 59} (1999) 064005; [\href{http://arxiv.org/abs/hep-th/9811056}{{\tt arXiv:hep-th/9811056}}].
		
		\bibitem{HReal}
		S.W. Hawking and H.S. Reall,
		Charged and rotating AdS black holes and their CFT duals,  {\it Phys.Rev.} {\bf D 61} (2000) 024014;  [\href{http://arxiv.org/abs/hep-th/9908109}{{\tt arXiv:hep-th/9908109}}].
		
		
  
\bibitem{NAS}
A. Nata Atmaja and K. Schalm,
Anisotropic Drag Force from 4D Kerr-AdS Black Holes,
{\it JHEP} {\bf 1104} (2011) 070; [\href{http://arxiv.org/abs/1012.3800}{{\tt arXiv:1012.3800 [hep-th]}}]

		
\bibitem{Bantilan:2018vjv}
  H.~Bantilan, T.~Ishii and P.~Romatschke, Holographic Heavy-Ion Collisions: Analytic Solutions with Longitudinal Flow, Elliptic Flow and Vorticity,
{\it Phys.\ Lett.}  B {\bf 785}, 201 (2018);  [\href{https://arxiv.org/abs/1803.10774}{{\tt arXiv:1803.10774[nucl-th]}}].

\bibitem{Kaminski}
M.~Garbiso and M.~Kaminski,
Hydrodynamics of simply spinning black holes \& hydrodynamics for spinning quantum fluids,
{\it JHEP} \textbf{12} (2020), 112;  [\href{https://arxiv.org/abs/2007.04345}{{\tt arXiv:2007.04345[hep-th]}}].

\bibitem{Arefeva:2020knc}
I.~Aref'eva, A.~Golubtsova and E.~Gourgoulhon, On the Drag Force of a Heavy Quark via 5d Kerr-AdS Background,
{\it Phys. Part. Nucl.} \textbf{51} (2020) no.4, 535-539.

\bibitem{AGG2020}
I. Ya. Aref'eva, A.A. Golubtsova and E. Gourgoulhon, Holographic drag force in 5d Kerr-AdS black hole, 
{\it JHEP} 04 (2021) 169  [\href{https://arxiv.org/abs/2004.12984}{{\tt arXiv:2004.12984[hep-th]}}].

\bibitem{GGU2021}
A.A. Golubtsova, E. Gourgoulhon and M.K. Usova, Heavy quarks in rotating plasma via holography,  {\it Nucl.Phys.} {\bf B} 979 (2022) 115786;
 [\href{https://arxiv.org/abs/2107.11672}{{\tt arXiv:2107.11672[hep-th]}}].


\bibitem{Petkou}
A. Armoni, J. L. F. Barbon, A. C. Petkou,
Orbiting strings in AdS black holes and N=4 SYM at finite temperature, {\it JHEP} 0206 (2002) 058; 
\href{http://arxiv.org/abs/hep-th/0205280}{{\tt arXiv:hep-th/0205280}}].


\bibitem{Alishahiha:2002fi}
M.~Alishahiha and A.~E.~Mosaffa,
Circular semiclassical string solutions on confining AdS / CFT backgrounds,
JHEP \textbf{10} (2002), 060; \href{http://arxiv.org/abs/hep-th/0210122}{{\tt arXiv:hep-th/0210122}}].

		
		\bibitem{Berman}
		D. S. Berman, M. K. Parikh, Holography and Rotating AdS Black Holes,
		{\it Phys.Lett.} {\bf B463} (1999) 168-173, [\href{http://arxiv.org/abs/hep-th/9907003}{{\tt  hep-th/9907003}}].
		
		
		\bibitem{AJ}
		A. M. Awad, C. V. Johnson,
		Higher dimensional Kerr-AdS black holes and the AdS/CFT correspondence,
		{\it Phys.Rev.} {\bf D 61} (2001) 124023;   	  [\href{http://arxiv.org/abs/hep-th/0008211}{{\tt arXiv:hep-th/0008211}}].	
		
		\bibitem{Gibbons:2004ai}
		G.~W.~Gibbons, M.~J.~Perry and C.~N.~Pope,
		The First law of thermodynamics for Kerr-anti-de Sitter black holes,
		{\it Class. Quant. Grav.}  {\bf 22}, 1503 (2005);  [\href{http://arxiv.org/abs/hep-th/0408217}{{\tt arXiv:hep-th/0408217}}].
		
		\bibitem{PaSk}
		I. Papadimitriou and K. Skenderis, Thermodynamics of Asymptotically Locally AdS Spacetimes,
		{\it JHEP} 0508 (2005)004 ; [\href{http://arxiv.org/abs/hep-th/0505190}{{\tt arXiv:hep-th/0505190}}].
		
		\bibitem{AAward:2007}
		A. M. Award,
		First Law, Counterterms and Kerr-AdS$_5$ Black Holes,
		{\it Int.J.Mod.Phys} {\bf D 18} (2007); [\href{http://arxiv.org/abs/0708.3458}{{\tt  arXiv:0708.3458 [hep-th]}}].
		
		\bibitem{MyPerry}
		R. C. Myers and M. J. Perry, Black Holes In Higher Dimensional Space-Times,{\it Annals. Phys.} {\bf 172} (1986)  304.
	
\bibitem{min} J. Minahan, Circular Semiclassical String Solutions on $AdS_5\times S_5$, Nucl.Phys. B {\bf 648} (2003) 203,[\href{http://arxiv.org/abs/hep-th/0209047}{{\tt  arXiv:hep-th/0209047}}].
%
\bibitem{Engquist:2003} J. Engquist, J. A. Minahan and K. Zarembo, Yang-Mills duals for semiclassical strings on $AdS_5\times S^5$, 
{\it JHEP} {\bf 11} (2003) 063, [\href{http://arxiv.org/abs/hep-th/0310188}{{\tt  arXiv:hep-th/0310188}}].
%
\bibitem{Dimov:2004} H. Dimov and R. C. Rashkov,  Generalized pulsating strings, 
{\it JHEP} {\bf 05} (2004) 068, [\href{http://arxiv.org/abs/hep-th/0404012}{{\tt  arXiv:hep-th/0404012}}].
%
\bibitem{Smedback:2004} M. Smedback,  Pulsating strings on $AdS_5\times S^5$,
{\it JHEP} 07 (2004) 004, [\href{http://arxiv.org/abs/hep-th/0405102}{{\tt  arXiv:hep-th/0405102}}].
%
\bibitem{Khan:2003sm}
A.~Khan, A.~L.~Larsen,Spinning pulsating string solitons in $AdS_5\times S^5$,
{\it Phys. Rev.}{\bf D} {\bf 69}, 026001 (2004), [\href{http://arxiv.org/abs/hep-th/0310019}{{\tt  arXiv:hep-th/0310019}}].

\bibitem{Arutyunov:2003za}
G.~Arutyunov, J.~Russo, A.~A.~Tseytlin, Spinning strings in $AdS_5\times S^5$: New integrable system relations,
Phys.\ Rev. D {\bf 69}, 086009 (2004),   [\href{http://arxiv.org/abs/hep-th/0311004}{{\tt  arXiv:hep-th/0311004}}].

\bibitem{Kruczenski:2004cn}
M.~Kruczenski and A.~A.~Tseytlin, Semiclassical relativistic strings in $S^5$ and long coherent operators  in $\mathcal{N} = 4$ SYM theory,
JHEP {\bf 0409}, 038 (2004), [\href{http://arxiv.org/abs/hep-th/0406189}{{\tt  arXiv:hep-th/0406189}}].

\bibitem{Bobev:2004id}
N.~P.~Bobev, H.~Dimov and R.~C.~Rashkov, Pulsating strings in warped $AdS_6 \times S^4$ geometry, [\href{http://arxiv.org/abs/hep-th/0410262}{{\tt  arXiv:hep-th/0410262}}].

\bibitem{Park:2005kt}
I.~Y.~Park, A.~Tirziu and A.~A.~Tseytlin, Semiclassical circular strings in $AdS_5$ and 'long' gauge field strength operators,
{\it Phys. Rev.} {\bf D} {\bf 71}, 126008 (2005),	 [\href{http://arxiv.org/abs/hep-th/0505130}{{\tt  arXiv:hep-th/0505130}}]. 

\bibitem{deVega:1994yz}
H.~J.~de Vega, A.~L.~Larsen, N.~G.~Sanchez, Semiclassical quantization of circular strings in de Sitter and anti-de Sitter space-times,
{\it Phys. Rev.}  {\bf D51}, 6917-6928 (1995),	\href{http://arxiv.org/abs/hep-th/9410219}{{\tt  arXiv:hep-th/9410219}}]. 
\bibitem{Chen:2008qq}
B.~Chen and J.~B.~Wu, Semi-classical strings in $AdS_4 \times CP^3$,
{\it JHEP} {\bf 0809}, 096 (2008),  [\href{http://arxiv.org/abs/0807.0802}{{\tt  arXiv:0807.0802 [hep-th]}}].	
\bibitem{Dimov:2009rd}
H.~Dimov and R.~C.~Rashkov, ``On the pulsating strings in $AdS_4 \times CP3$,''
{\it Adv. High Energy Phys.} {\bf 2009}, 953987 (2009), [\href{http://arxiv.org/abs/0908.2218}{{\tt  arXiv:0908.2218 [hep-th]}}].

\bibitem{Arnaudov:2010by}
D.~Arnaudov, H.~Dimov and R.~C.~Rashkov, On the pulsating strings in $AdS_5 \times T^{1,1}$,
{\it J. Phys.} {\bf A} {\bf 44}, 495401 (2011),	  [\href{http://arxiv.org/abs/1006.1539}{{\tt  arXiv:1006.1539 [hep-th]}}]. 

\bibitem{Arnaudov:2010dk}
D.~Arnaudov, H.~Dimov and R.~C.~Rashkov, On the pulsating strings in Sasaki-Einstein spaces,
{\it AIP Conf. Proc.}  {\bf 1301}, 51 (2010),   [\href{http://arxiv.org/abs/1007.3364}{{\tt  arXiv:1007.3364 [hep-th]}}]. 

\bibitem{Beccaria:2010zn}
M.~Beccaria, G.~V.~Dunne, G.~Macorini, A.~Tirziu and A.~A.~Tseytlin, Exact computation of one-loop correction to energy of pulsating strings in $AdS_5 \times S^5$,
{\it J. Phys. } {\bf A} {\bf 44}, 015404 (2011), [\href{http://arxiv.org/abs/1009.2318}{{\tt  arXiv:1009.2318 [hep-th]}}].  

\bibitem{Giardino:2011jy}
S.~Giardino and V.~O.~Rivelles, Pulsating Strings in Lunin-Maldacena Backgrounds,
{\it JHEP} {\bf 1107}, 057 (2011),  [\href{http://arxiv.org/abs/1105.1353}{{\tt  arXiv:1105.1353 [hep-th]}}]. 

\bibitem{Pradhan:2013sja}
P.~M.~Pradhan and K.~L.~Panigrahi, Pulsating Strings With Angular Momenta,
{\it Phys. Rev.} {\bf D} {\bf 88}, 086005 (2013), [\href{http://arxiv.org/abs/1306.0457}{{\tt  arXiv:1306.0457 [hep-th]}}].  

\bibitem{Pradhan:2014zqa}
P.~M.~Pradhan, Oscillating Strings and Non-Abelian T-dual Klebanov-Witten Background,
 {\it Phys. Rev.} {\bf D }{\bf 90}, 046003 (2014),	[\href{http://arxiv.org/abs/1406.2152}{{\tt  arXiv:1406.2152 [hep-th]}}].
%
%

\bibitem{Kraus:2017kyl}
P.~Kraus, A.~Sivaramakrishnan and R.~Snively, Black holes from CFT: Universality of correlators at large c,
{\it JHEP} \textbf{08} (2017), 084, 	[\href{http://arxiv.org/abs/1706.00771}{{\tt  arXiv:1706.00771 [hep-th]}}].


\bibitem{Iliesiu:2018fao}
L.~Iliesiu, M.~Kolo\u{g}lu, R.~Mahajan, E.~Perlmutter and D.~Simmons-Duffin, The Conformal Bootstrap at Finite Temperature,
{\it JHEP} \textbf{10} (2018), 070, [\href{http://arxiv.org/abs/1802.10266}{{\tt  arXiv:1802.10266 [hep-th]}}].


\bibitem{Kulaxizi:2018dxo}
M.~Kulaxizi, G.~S.~Ng and A.~Parnachev, Black Holes, Heavy States, Phase Shift and Anomalous Dimensions,
{\it SciPost Phys.} \textbf{6} (2019) no.6, 065,  [\href{http://arxiv.org/abs/1812.03120}{{\tt  arXiv:1812.03120 [hep-th]}}].


\bibitem{Fitzpatrick:2019zqz}
A.~L.~Fitzpatrick and K.~W.~Huang, Universal Lowest-Twist in CFTs from Holography,
{\it JHEP} \textbf{08} (2019), 138,  [\href{http://arxiv.org/abs/1903.05306}{{\tt  arXiv:1903.05306 [hep-th]}}].


\bibitem{Karlsson:2019qfi}
R.~Karlsson, M.~Kulaxizi, A.~Parnachev and P.~Tadi\'c, Black Holes and Conformal Regge Bootstrap,
{\it JHEP} \textbf{10} (2019), 046, [\href{http://arxiv.org/abs/1904.00060}{{\tt  arXiv:1904.00060 [hep-th]}}].


\bibitem{Kulaxizi:2019tkd}
M.~Kulaxizi, G.~S.~Ng and A.~Parnachev, Subleading Eikonal, AdS/CFT and Double Stress Tensors,
{\it  JHEP} \textbf{10} (2019), 107, [\href{http://arxiv.org/abs/1907.00867}{{\tt  arXiv:1907.00867 [hep-th]}}].

\bibitem{Alday:2020eua}
L.~F.~Alday, M.~Kologlu and A.~Zhiboedov, Holographic Correlators at Finite Temperature,
 [\href{http://arxiv.org/abs/2009.10062}{{\tt  arXiv:2009.10062 [hep-th]}}].


\bibitem{Grinberg:2020fdj}
M.~Grinberg and J.~Maldacena, Proper time to the black hole singularity from thermal one-point functions,
{\it JHEP} \textbf{03} (2021), 131, [\href{http://arxiv.org/abs/2011.01004}{{\tt  arXiv:2011.01004 [hep-th]}}].


\bibitem{Parnachev:2020zbr}
A.~Parnachev and K.~Sen, Notes on AdS-Schwarzschild eikonal phase,
{\it JHEP} \textbf{03} (2021), 289, [\href{http://arxiv.org/abs/2011.06920}{{\tt  arXiv:2011.06920 [hep-th]}}].

	
\bibitem{Karlsson:2021duj}
R.~Karlsson, A.~Parnachev and P.~Tadi\'c, Thermalization in Large-N CFTs, [\href{http://arxiv.org/abs/2102.04953}{{\tt  arXiv:2102.04953 [hep-th]}}].


\bibitem{Rodriguez-Gomez:2021pfh}
D.~Rodriguez-Gomez and J.~G.~Russo, Correlation functions in finite temperature CFT and black hole singularities,
{\it JHEP} \textbf{06} (2021), 048, [\href{http://arxiv.org/abs/2102.11891}{{\tt  arXiv:2102.11891 [hep-th]}}].


\bibitem{Litvinov:2013sxa} 
A.~Litvinov, S.~Lukyanov, N.~Nekrasov and A.~Zamolodchikov,  Classical Conformal Blocks and Painleve VI,
{\it JHEP} \textbf{07} (2014), 144,  [\href{http://arxiv.org/abs/1309.4700}{{\tt  arXiv:1309.4700 [hep-th]}}]. 

\bibitem{Novaes:2014lha}
F.~Novaes and B.~Carneiro da Cunha, Isomonodromy, Painlev\'e transcendents and scattering off of black holes,
{\it JHEP}  \textbf{07} (2014), 132, [\href{http://arxiv.org/abs/1404.5188}{{\tt  arXiv:1404.5188 [hep-th]}}].

\bibitem{Amado:2017kao}
J.~B.~Amado, B.~Carneiro da Cunha and E.~Pallante, On the Kerr-AdS/CFT correspondence,
JHEP \textbf{08} (2017), 094, [\href{http://arxiv.org/abs/1702.01016}{{\tt  arXiv:1702.01016 [hep-th]}}].

\bibitem{BarraganAmado:2018zpa}
J.~Barrag\'an Amado, B.~Carneiro Da Cunha and E.~Pallante,  Scalar quasinormal modes of Kerr-AdS${_5}$,
{\it Phys. Rev.} {\bf D} \textbf{99} (2019) no.10, 105006, [\href{http://arxiv.org/abs/1812.08921}{{\tt  arXiv:1812.08921 [hep-th]}}].

\bibitem{Amado:2020zsr}
J.~B.~Amado, B.~Carneiro da Cunha and E.~Pallante, Vector perturbations of Kerr-AdS$_{5}$ and the Painlev\'e VI transcendent,
{\it JHEP} \textbf{04} (2020), 155,  [\href{http://arxiv.org/abs/2002.06108}{{\tt  arXiv:2002.06108 [hep-th]}}]. 

\bibitem{BarraganAmado:2021uyw}
J.~Barrag\'an Amado, B.~Carneiro da Cunha and E.~Pallante, Remarks on holographic models of the Kerr-AdS$_{5}$ geometry,''
{\it JHEP} \textbf{05} (2021), 251,  [\href{http://arxiv.org/abs/2102.02657}{{\tt  arXiv:2102.02657 [hep-th]}}].	
		
	\end{thebibliography}
\end{document}